\documentclass[lettersize,journal]{IEEEtran}

\usepackage{amsmath,amsfonts}
\usepackage{algorithmic}
\usepackage{algorithm}
\usepackage{array}
\usepackage[caption=false,font=footnotesize,labelfont=sf,textfont=sf]{subfig}
\usepackage{textcomp}
\usepackage{stfloats}
\usepackage{url}
\usepackage{verbatim}
\usepackage{graphicx}
\usepackage{cite}
\usepackage{multirow}
\usepackage[colorlinks,linkcolor=black,anchorcolor=black,citecolor=black]{hyperref}
\begin{document}

\title{KCES: A Workflow Containerization Scheduling Scheme Under Cloud-Edge Collaboration Framework}

\author{Chenggang Shan, Runze Gao, Qinghua Han, Zhen Yang, 
        Jinhui Zhang, and Yuanqing Xia,~\IEEEmembership{Fellow,~IEEE}
        
\IEEEcompsocitemizethanks{
  \IEEEcompsocthanksitem C. Shan, Q. Han, and Z. Yang are with the School of Artificial Intelligence, Zaozhuang University, Zaozhuang, Shandong, 277160, China. E-mail:~\{uzz\_scg, hanqinghua123, yang\_zh99\}@163.com 
  \IEEEcompsocthanksitem R. Gao, J, Zhang, and Y. Xia are with the School of Automation, Beijing institute of technology, Beijing, 10081, China. E-mail: \{runze\_gao, zhangjinh, xia\_yuanqing\}@bit.edu.cn
}
\thanks{
(Corresponding authors: Zhen Yang; Jinhui Zhang; Yuanqing Xia)}
}
\markboth{}%
{C. Shan \MakeLowercase{\textit{et al.}}: KCES: A Workflow Containerization Scheduling Scheme Under Cloud-Edge Collaboration Framework}


\maketitle

\begin{abstract}
As more IoT applications gradually move towards the cloud-edge collaborative mode, the containerized scheduling of workflows extends from the cloud to the edge. However, given the high delay of the communication network, loose coupling of structure, and resource heterogeneity between cloud and edge, workflow containerization scheduling in the cloud-edge scenarios faces the difficulty of resource coordination and application collaboration management.
To address these two issues, we propose a KubeEdge-Cloud-Edge-Scheduling scheme named KCES, a workflow containerization scheduling scheme for the KubeEdge cloud-edge framework. The KCES includes a cloud-edge workflow scheduling engine for KubeEdge and workflow scheduling strategies for task horizontal roaming and vertical offloading. Considering the scheduling optimization of cloud-edge workflows, this paper proposes a cloud-edge workflow scheduling model and cloud-edge node model and designs a cloud-edge workflow scheduling engine to maximize cloud-edge resource utilization under the constraint of workflow task delay. 
A cloud-edge resource hybrid management technology is used to design the cloud-edge resource evaluation and resource allocation algorithms to achieve cloud-edge resource collaboration. Based on the ideas of distributed functional roles and the hierarchical division of computing power, the horizontal roaming among the edges and vertical offloading strategies between the cloud and edges for workflow tasks are designed to realize the cloud-edge application collaboration. 
Through a customized IoT application workflow instance, experimental results show that KCES is superior to the baseline in total workflow duration, average workflow duration, and resource usage and has the capabilities of horizontal roaming and vertical offloading for workflow tasks.
\end{abstract}


\section{Introduction}
\IEEEPARstart{W}{ith} the rapid development of the Internet of Thing~(IoT) and artificial intelligence technology, the massive data created by vast access devices is exponential growth, putting enormous pressure on network bandwidth and servers~\cite{xia2015cloud}. Traditional cloud computing gradually presents problems such as poor real-time performance, insufficient bandwidth, high energy consumption, and lack of data security~\cite{varghese2016challenges,peng2021constrained}. To this end, edge computing, as a new computing model, provides an edge-assisted cloud way to cope with the shortcomings of cloud computing. 
It provides a cloud-edge collaborative distributed computing architecture closer to data sources, integrated computing, network, storage, and applications, and meets key needs of agile link, real-time business, application intelligence, and security and privacy protection in the IoT field through near-device service mode~\cite{shi2016edge}.

Cloud workflow scheduling is a significant means to realize the efficient execution of workflow on the cloud platform. It achieves the best resource match between workflow tasks and cloud computing resources by optimizing scheduling~\cite{yang2023look}. In the cloud-native era, container technology facilitates the gradual extension of cloud workflow scheduling from the cloud to the edge~\cite{shan2021containerized,shan2023kubeadaptor,shan2023adaptive}.
In addition to dealing with the processing of massive data of edge devices, the distributed computing architecture of cloud-edge collaboration can release the execution ability of cloud workflows to the edge~\cite{haja2019sharpening} and meet the practical needs of workflow containerization applications in the IoT and industrial internet field. However, because of the high time delay, loose coupling structure, and resource heterogeneity of the communication network between cloud and edges, workflow containerization scheduling in cloud-edge collaborative architecture faces two challenges. This paper explains how we address these two challenges and comprehensively describes our solution. 

One notable challenge is the difficulty of collaborative management of cloud-edge resources. The loose coupling structure and high time latency of the cloud-edge network make the workflow containerization scheduling under the cloud-edge framework unable to obtain the cloud-edge resource utilization rate in time and delay the best time of task optimization schedule \cite{bader2021tarema,agrawal2019iot}.
Another challenge is the difficulty of application collaboration management. Cloud-edge application collaboration enables cloud and edge to host workflow task containers through workflow containerization scheduling, realizes vertical offloading between cloud and edges and horizontal roaming among edges for workflow tasks, and completes workflow containerization execution under task dependency and delay constraints. 
However, the high time delay of the communication network between cloud and edges restricts the real-time communication of cloud-edge hosting tasks, and resource heterogeneity makes the same task depend on different container images on cloud and edge nodes, along with loose coupling structure, which affects the roaming and offloading of cloud-edge tasks and poses challenges to application collaboration management~\cite{hao2019itaskoffloading, zhang2019collaborative}.

At present, related solutions like KubeEdge~\cite{xiong2018extend,kubeedge2023} from Huawei and OpenYurt~\cite{openyurt2023} from Alibaba have extended the Kubernetes (K8s) container orchestration capabilities to the edge. In addition, the proposed solutions for K8s, such as Cloud4IoT~\cite{dupont2017edge} and FLEDGE~\cite{goethals2019fledge}, solve the horizontal roaming and vertical offloading problems of IoT applications, and the edge container orchestration and scheduling problems, respectively. 
The cloud-edge collaboration framework proposed by the industry integrates cloud-edge heterogeneous resource management and focuses on industry applications, lacking efficient scheduling strategies for request assignment and service orchestration. 
The academic pays more attention to task scheduling and deployment rather than workflow scheduling research in the IoT field, and there is a lack of research on cloud workflow scheduling optimization in real cloud-edge collaboration scenarios.

To overcome these challenges, we propose a workflow containerization scheduling scheme for the cloud-edge framework named KCES in this paper. 
The KCES includes a cloud-edge workflow scheduling engine for KubeEdge and workflow scheduling strategies for task horizontal roaming and vertical offloading in the cloud workflow scheduling model. Considering the scheduling optimization of cloud-edge workflows, this paper proposes a cloud-edge workflow scheduling model and cloud-edge node model and designs a cloud-edge workflow scheduling engine to maximize cloud-edge resource utilization under the constraint of workflow task delay.
The KCES employs resource hybrid management technology for cloud-edge nodes to design the cloud-edge resource evaluation and resource allocation algorithms to achieve cloud-edge resource collaboration. Based on the ideas of distributed functional roles and the hierarchical division of computing power, the horizontal roaming among the edges and vertical offloading strategies between the cloud and edges for workflow tasks are designed to realize the cloud-edge application collaboration.

This paper aims to open up the application collaboration and resource collaboration between cloud and edges and realize task horizontal roaming among edges and vertical offloading between cloud and edges under the workflow scheduling model. Compared with the baseline algorithm, experimental results under three distinct workflow arrival patterns show that our KCES gains time-saving of 8\% to 24.5\% in the average total duration of all workflows, time-saving of 22.7\% to 44.5\% in the average duration of individual workflow, and an increase of 20\% and 6.6\% to 13.3\% in CPU and memory resource usage rate, respectively. Our contributions are summarized as follows:
\begin{enumerate}
        \item \textbf{Model:} We propose a workflow scheduling model for cloud-edge scenarios and build a workflow scheduling optimization problem in pursuit of cloud-edge resource utilization maximization.
        \item \textbf{Platform:} We design a cloud-edge workflow scheduling engine under the KubeEdge cloud-edge framework and realize cloud-edge containerized scheduling under the constraints of workflow task dependency.
        \item \textbf{Algorithms:} A resource hybrid management technology for cloud-edge nodes is used to design the resource evaluation and resource allocation algorithm to achieve cloud-edge resource collaboration. Based on the idea of distributed functional roles and hierarchical division of computing power, the horizontal roaming among the edges and vertical offloading strategies between the cloud and edges of cloud workflow tasks are designed to realize the cloud-edge application collaboration.
        \item \textbf{Evaluation:} We customize a universal workflow instance of IoT application and verify the containerized scheduling capability of cloud-edge workflow scheduling engine within the KCES and the superior performance of task horizontal roaming and vertical offloading algorithms.
\end{enumerate}

The rest of the paper is organized as follows. Section 2 summarizes the
related work. Section 3 presents details of the system model and formulation. Section 4 elaborates on the cloud-edge workflow scheduling engine and its modular description. Section 5 further introduces algorithm implementation. Section 6 provides the experimental analysis and evaluates the effectiveness of the KCES. Finally, we conclude the paper in Section 7. 

We have open-sourced the KCES. The source code is publicly available on GitHub~\cite{experiment2023kCES}.

\section{Related Work}
This section analyzes the research progress related to cloud-edge collaboration from the perspectives of industry and academia, then introduces containerized workflow scheduling and presents a novelty analysis of our work.

\subsection{Cloud-Edge Collaboration}
The collaboration between cloud and edge is the most explored collaboration form by researchers and industry at present, and it is also a relatively mature collaborative model in the technical stage. In industry, some companies have released products based on cloud-edge collaboration. KubeEdge~\cite{kubeedge2023} was first proposed by Huawei as an open-source intelligent edge framework for cloud-edge collaboration. Built on K8s, KubeEdge supports cloud-native edge computing and extends containerized application orchestration capabilities to the edge~\cite{xiong2018extend}. Alibaba contributed the complete edge computing cloud-native project OpenYurt to the community~\cite{openyurt2023}. Adhering to the concept of cloud-edge integration, OpenYurt relies on the powerful container arrangement and scheduling capabilities of native K8s and numerous edge computing application scenarios and implements a complete set of edge cloud-native solutions. 
Baetyl~\cite{baetyl2023} was an edge computing framework released by Baidu, which has the characteristics of containerization of component modules and one-click deployment. It can seamlessly extend cloud computing, data, and service to edge devices, provide temporary offline and low-latency computing services, and meet various edge computing scenarios. 
These frameworks help users solve the issues of large-scale application delivery, operation and maintenance, and control on massive edge and end resources and also provide a sinking channel for central services to achieve seamless docking with edge computing applications.
The cloud-edge collaboration framework proposed by the industry pays more attention to heterogeneous resource management, application delivery, operation, and maintenance in the cloud-edge scenario rather than workflow scheduling, so it lacks an efficient scheduling framework for workflow task assignment and cloud-edge service orchestration.


The academic community has also made corresponding advances for cloud-edge collaboration frameworks. 
Haja et al.~\cite{haja2019sharpening} designed a K8s scheduler with edge topology awareness and self-healing characteristics to deal with the deployment of delay-sensitive applications at the edge while considering the delay constraints and edge reliability of applications. 
Huang et al.~\cite{huang2019design} proposed a machine learning framework for training in the cloud and prediction at the edge and uses Docker and K8s cloud-native tools to build an edge machine learning platform. 
In order to solve the problem of edge data confidentiality, Dupont et al.~\cite{dupont2017edge} proposed Cloud4IoT, a platform for K8s to perform horizontal roaming and vertical offloading of IoT application tasks. 
In~\cite{goethals2019fledge}, Goethals et al. proposed the edge container orchestration tool FLEDGE based on K8s and studied the container orchestration technology, container runtime selection, and container network implementation under limited edge resources. 
In~\cite{wang2020kubeedge}, Wang et al. proposed a KubeEdge-based edge AI framework that helps reduce the burden of developing specific edge/embedded AI systems and promotes cloud-edge coordination and collaboration. 
Considering request assignment and service orchestration, Han et al.~\cite{han2021tailored} proposed a K8s-oriented KaiS scheduling framework for an edge-cloud system, a learning-based scheduling framework to satisfy service requests using distributed edge nodes and cloud cluster resources. 
However, these frameworks involve lightweight container technology, suitable for the K8s platform but not for workflow application scenarios with dependencies. 

\subsection{Containerized Workflow Scheduling}
Recently, containerized workflow scheduling has emerged by virtue of the rapid deployment and resource-sharing characteristics of container technology. Extended to the field of cloud-edge collaboration, Li et al.~\cite{li2022wholistic} proposed a three-step scheduling model combining flexible resource allocation and optimization deployment for containers under workflow constraints. The authors considered multiple objectives of cloud-edge resources and containerized workflows, including makespan, load imbalance, and energy consumption. Meng et al.~\cite{meng2019online} put forth an online algorithm named Dedas that greedily schedules newly arriving tasks and considers whether to replace some existing tasks to meet the deadline of the arriving task as much as possible. In \cite{qian2020workflow}, Qian et al. proposed a workflow-aided IoT paradigm with intelligent edge computing to automate the execution of IoT applications with dependencies, aiming to reduce the latency of the IoT systems. For cloud-native workflow scheduling, Shin et al.~\cite{shin2022cloud} proposed a workflow-aware scheduling algorithm with a hybrid priority rule and dynamic task parallelism to improve resource utilization and minimize weighted workflow completion time. 
These research works pay more attention to the optimal scheduling and deployment of tasks, aiming to improve various indicators of workflow, including completion time, energy consumption, and resource utilization. The verification platforms tend to be simulation testbeds and customized testbeds and do not rely on open-source edge computing frameworks, which is slightly insufficient in practice and application promotion.

\subsection{Novelty Analysis of Our Work}
Different from the above research, our proposed scheme KCES makes full use of the technical advantages and community influence of K8s and KubeEdge and realizes the containerized execution of IoT workflows in a real-world cloud-edge environment with the cloud-edge workflow scheduling engine and workflow scheduling scheme for KubeEdge. 
The KCES solves the issues of cloud-edge resource collaboration and application collaboration and maximizes cloud-edge resource usage under workflow constraints. It equips with the scheduling ability of task horizontal roaming and vertical offloading to achieve load balancing in cloud-edge clusters. 

\section{System Model and Problem Formulation}
This section describes the workflow scheduling model in the cloud-edge scenario, presents the cloud and edge node model, and formulates workflow scheduling optimization problems for cloud-edge scenarios under the task delay constraint.

\subsection{Workflow Scheduling Model}
There is a KubeEdge cluster in a cloud-edge scenario, where the master node does not participate in the load (only manages the cluster). $C = \{C_1, C_2,..., C_m\}$ stands for a group of cloud virtual machine nodes and $m$ represents the number of cloud nodes in the cluster. For $m$ cloud nodes, there is a set of available CPU cores $\{u^{c}_1, u^{c}_2,..., u^{c}_m\}$ and a set of available memory capacities $\{mem^{c}_1,mem^{c}_2,...,mem^{c}_m\}$ to represent its available resources. Another group of edge nodes in the cluster is represented as $E=\{E_1, E_2,..., E_n\}$, $n$ indicates the number of edge nodes in the cluster. For $n$ edge nodes, there is a set of available CPU cores $\{u^{e}_1, u^{e}_2,..., u^{e}_n\}$and a set of available memory capacities $\{mem^{e}_1,mem^{e}_2,...,mem^{e}_n\}$ to represent its available resources.

A workflow set in the cloud-edge scenario is expressed as $W=\{w_1,w_2,...,w_k\}$, where $k$ is the number of workflows. A workflow abstraction is defined as $w_i = \{sla_{w_i}, s_{i,1}, s_{i,2},...,s_{i,n}\}$, where $i$ represents the sequence number of the workflow, $sla_{w_i}$ represents the Service Level Agreement~(SLA) of the workflow, $s_{i,n}$ indicates a step or task in workflow $w_i$. Each workflow task is denoted as: 
\begin{equation}
  \label{eq_s_ij}
  \begin{split}
s_{i,j}=\{sla_{s_{i,j}},id,data,image,cpu,mem,duration\},\\~1\leq i \leq k~and~1\leq j \leq n.   
  \end{split}
\end{equation}

The $id$ of the above (\ref{eq_s_ij}) indicates the unique identifier of the task in the workflow $w_i$, $data$ indicates the needed data during the task execution process, and $image$ indicates the image address of the workflow task. $cpu$ and $mem$ denote the number of CPU millicore and memory capacity (Mi) required for this task, respectively. The $duration$ indicates the duration of the task Pod\footnote[1]{Pod is an atomic scheduling unit of the K8s cluster that can encapsulate one or more task containers.}. Each workflow has an optional SLA ($sla_i$), either on the workflow $sla_{w_i}$ or the workflow task $sla_{s_{i,j}}$, consisted of $slo_1, slo_2,... slo_n$ service level objectives as follows:
\begin{eqnarray}
  \label{eq_sla_i}
sla_{i}=\{slo_{1},slo_{2},...,slo_{n}\},i\in\{w_{i},s_{i,j}\}.
\end{eqnarray}

The cloud-edge scenario only considers the deadline as a single Service Level Object~(SLO), and each task in the workflow needs to be completed within its deadline. Likewise, the workflow is no exception. Since the edge scenario is a hard real-time system in which the deadline of each task and workflow should consider the real-time requirements of the terminal equipment in the edge scenario.
\begin{eqnarray}
  \label{eq_sla}
  \begin{split}
  &sla_{w_i}=deadline_{w_i},\\
  &sla_{s_{i,j}}=deadline_{s_{i,j}}.
  \end{split}
\end{eqnarray}

Noted that the last task in the workflow, $s_{i,last}$, has the same deadline as this workflow deadline:
\begin{equation}
  \label{eq_deadline}
  deadline_{s_{i,last}}=deadline_{w_i}.
\end{equation}

For intensive diagnostic workflow in IoT scenarios, delay-sensitive tasks and tasks with data privacy need to be scheduled by the cloud-edge workflow scheduling engine to the edge node near the device end. If the edge node cannot complete the task on time, it means that this edge node does not have enough resources to perform the task. The cloud-edge workflow scheduling engine will notify the user to reset the task deadline, run the node model at the edge again, or select another edge node in the same edge scenario.

\subsection{Edge Node Model}
If a task is assigned to an edge node, the cloud-edge workflow scheduling engine evaluates whether the edge node can run this task based on the task's execution time.
The execution time of a task on the edge node includes the computing time, data transmission time from the device to the edge node, and task image transmission time, which are $T^{e,com}_{s_{i,j}}$, $T^{e,data}_{s_{i,j}}$, and $T^{e,img}_{s_{i,j}}$, respectively. Assuming that the amount of task output data is small, the transmission delay of output data is negligible. The execution time of task $s_{i,j}$ of workflow $w_i$ on edge node $e$ is:
\begin{equation}
  \label{eq_edge_exec_time} 
T^{e}_{s_{i,j}}=T^{e,com}_{s_{i,j}}+T^{e,data}_{s_{i,j}}+T^{e,img}_{s_{i,j}}.
\end{equation}
The calculation time of workflow task $s_{i,j}$ at edge node $e$ in the above~(\ref{eq_edge_exec_time}) can be obtained by the following equation:
\begin{equation}
  \label{eq_edge_computing_time} 
T^{e,com}_{s_{i,j}}=V^{data}_{s_{i,j}}/R^{allocated,e}_{s_{i,j},cpu},
\end{equation}
$V^{data}_{s_{i,j}}$ indicates the data amount collected by the device end that needs to be processed on edge node $e$ for workflow task $s_{i,j}$. $R^{allocated,e}_{s_{i,j},cpu}$ represents the computing resources allocated by edge node $e$ within the lifecycle of task $s_{i,j}$ (only CPU is considered). The resource requirement for each workflow task can be expressed as:
\begin{equation}
  \label{eq_wf_task_resource_requirement} 
Re^{cpu}_{s_{i,j}}=\lambda^{e}_{s_{i,j}}\cdot V^{data}_{s_{i,j}},
\end{equation}
where $\lambda^{e}_{s_{i,j}}$ represents the number of instructions (MIPS) required to process each byte of workflow task $s_{i,j}$ image data on edge node $e$, which can be measured offline~\cite{miettinen2010energy}. To ensure that the tasks scheduled to the edge node are completed on time (meeting the SLA), the resources allocated to this task by the edge node $e$ must be no less than the required resources, as follows:
\begin{equation}
  \label{eq_wf_task_allocated_resource_condition} 
Re^{cpu}_{s_{i,j}}\leq R^{allocated,e}_{s_{i,j},cpu}.
\end{equation}

The workflow task $s_{i,j}$ scheduled on the edge node requires the device connected to the edge node to transmit the collected data to the edge node, and the data transmission time can be calculated as follows:
\begin{equation}
  \label{eq_data_transmit_time_from_device_to_edge} 
T^{e,data}_{s_{i,j}}=V^{data}_{s_{i,j}}/W_e,
\end{equation}
where ${W_e}^{-1}$ indicates the time required for a byte of data to be transmitted over the communication link between the edge node and the device.

Considering the limited storage resources of edge nodes, the $Image_{s_{i,j}}$ of workflow task $s_{i,j}$ is only partially stored in the local edge node, and the transfer time of workflow task image can be calculated as follows:
\begin{eqnarray}
  \label{eq_img_tran_time} 
T_{s_{i,j}}^{e,img}=
\left\{\begin{array}{cc}
0,~~~~~~~~~~~~~~&\mbox{$Image_{s_{i,j}}\in I^{e}$}\,\,\\
V_{s_{i,j}}^{img}/W_{e,c}^{s_{i,j}},~~~~~~~&\mbox{$Image_{s_{i,j}}\notin I^{e}$}
\end{array}
\right.
\end{eqnarray}

Herein, $I_e$ indicates the local image storage of edge node $e$, $V_{s_{i,j}}^{img}$ denotes the amount of image data for the workflow task $s_{i,j}$. $W_{e,c}^{s_{i,j}}$ indicates the link bandwidth allocated by the edge node to connect to the cloud node for task $s_{i,j}$.

In addition, the constraint of the edge node model is
\begin{equation}
  \label{eq_edge_exec_model_condition} 
\sum\limits_{{s_{i,j}}\in S_{p}^{e}}Re_{s_{i,j}} \leq R^{e},
\end{equation}
where $S_{p}^{e}$ represents the set of active workflow tasks on edge node at time $p$. This constraint means that the sum of resource requirements of tasks $s_{i,j}$ running on edge node $e$ at time $p$ is not greater than the resource quota $R^e$ of the edge node $e$.

\subsection{Cloud Node Model}
Non-latency sensitive tasks and non-data privacy tasks are scheduled by the workflow scheduling engine to the cloud node $c$. Since the workflow image warehouse is located in the cloud, we can ignore the task image transfer time. The execution time of workflow task $s_{i,j}$ on cloud node $c$ is calculated as follows:
\begin{eqnarray}
  \label{eq_cloud_exec_time}
  T_{s_{i,j}}^{c}=T_{s_{i,j}}^{c,com}+T_{s_{i,j}}^{e,data}+T_{s_{i,j}}^{e,c,data}.
\end{eqnarray}
$T_{s_{i,j}}^{c,com}$ represents the computing time of workflow task $s_{i,j}$ on cloud node $c$, which is calculated as follows:
\begin{eqnarray}
  \label{eq_cloud_computing_time}
  T_{s_{i,j}}^{c,com}=V_{s_{i,j}}^{data}/R_{s_{i,j},cpu}^{allocated,c},
\end{eqnarray}
$R_{s_{i,j},cpu}^{allocated,c}$ indicates the computing resources allocated by the cloud node within the lifecycle of task $s_{i,j}$ (only considering CPU).

The workflow task $s_{i,j}$ scheduled on the cloud node transmits the collected data from the device to the cloud node through the edge node. The data transmission time $T_{s_{i,j}}^{e,data}$ from the device to the edge node can be calculated by (\ref{eq_data_transmit_time_from_device_to_edge}). 

The transmission time of the collected data by the device from the edge node to the cloud node required for workflow task $s_{i,j}$ on the cloud node is calculated as follows: 
\begin{eqnarray}
  \label{eq_data_transmit_time_from_edge_to_cloud}
  T_{s_{i,j}}^{e,c,data}=V_{s_{i,j}}^{data}/W_{e,c}^{s_{i,j}}.
\end{eqnarray}

Delay constraints under both the edge node model and cloud node model satisfy:
\begin{eqnarray}
  \label{eq_edge_cloud_delay_constraint}
  \alpha_{s_{i,j}}^{e}T_{s_{i,j}}^{e}+\alpha_{s_{i,j}}^{c}T_{s_{i,j}}^c\leq deadline_{s_{i,j}},
\end{eqnarray}

In the above formula, $\alpha_{s_{i,j}}^{e}$ and $\alpha_{s_{i,j}}^{c}$ are binary variables, indicating whether workflow task $s_{i,j}$ is scheduled to the edge node or cloud node. If task $s_{i,j}$ runs on edge node $e$ or cloud node $c$, the corresponding binary variable is 1. Otherwise, it is 0. The binary variables must meet $\alpha_{s_{i,j}}^{e} \neq \alpha_{s_{i,j}}^{c}$.

\subsection{Problem Formulation}
This paper studies containerized scheduling of cloud workflow in the cloud-edge scenario, aiming to solve the issues of resource collaboration and application collaboration for workflow tasks in the cloud-edge scenario. Cloud-edge workflow scheduling optimization aims to host as many workflow tasks as possible while maximizing the utilization of cloud-edge resources. 

Whether in the edge node model or the cloud node model, once the workflow task goes through Resource Evaluator, Resource Allocator, and Scheduling Decision~(\ref{sec:cwb}), the Resource Manager immediately allocates the resources of the corresponding node to this task, and the task scheduling is done by the default scheduler Kube-scheduler~\footnote[2]{Kube-scheduler is the default scheduler for K8s and runs as part of the control plane.} of K8s. 
The workflow tasks in this paper are complete and uninterruptible and do not support the resource reallocation scheme during the lifecycle of task containers.
 We assume that each node in the cloud-edge cluster is always active, and the workflow continuously arrives at the cloud-edge workflow scheduling engine.
Since the CPU is a compressible resource and memory is an incompressible resource, the optimization goal of resource allocation only considers memory resources. Regardless of the current edge node or cloud node, the Resource Manager's optimization problem for node resource allocation at the arrival time of task $s_{i,j}$ is represented by
\begin{eqnarray}
  \label{eq_optimization_problem}
  \begin{split}
\operatorname{\min}\quad-\sum_{i=1}^{k}\sum_{j=1}^{n}\sum_{s_{i,j}\in S} R_{s_{i,j},mem}^{allocated} \big /mem,\\mem\in \left \{mem_p^c,mem_l^e  \right \},\\
  \end{split}
\end{eqnarray}
satisfying:
\begin{equation}
  \label{eq_optimization_problem_constraits}
  \begin{split}
&C_1:~\sum_{i=1}^{k}\sum_{j=1}^{n}\sum_{s_{i,j}\in S^p \cap S^r}R_{s_{i,j},cpu}^{allocated}\leq u,u\in \left \{u_p^c,u_l^e  \right \},\\
&C_2:~\sum_{i=1}^{k}\sum_{j=1}^{n}\sum_{s_{i,j}\in S^p \cap S^r}R_{s_{i,j},mem}^{allocated}\leq mem,\\&~~~~~~~mem\in \left \{mem_p^c,mem_l^e  \right \},\\
&C_3:~\alpha_{s_{i,j}}^{c}T_{s_{i,j}}^{c}+\alpha_{s_{i,j}}^{e}T_{s_{i,j}}^e\leq deadline_{s_{i,j}},\\
&~~~~~~~\alpha_{s_{i,j}}^c,\alpha_{s_{i,j}}^e \in \left\{0,1\right\}~and~\alpha_{s_{i,j}}^c\neq\alpha_{s_{i,j}}^e,\\
 \end{split}
\end{equation}

The Resource Manager allocates resources $R_{s_{i,j},cpu}^{allocated}$ and $R_{s_{i,j},mem}^{allocated}$ when the task $s_{i,j}$ arrives. The optimization problem of (\ref{eq_optimization_problem}) is to maximize the resource allocation of all task set $S$ contained by the current node (cloud node or edge node). Constraint $C_1$ means that at the arrival of the current task $s_{i,j}$, the total CPU resource allocation of all tasks hosted by the destination node cannot exceed resource quota $u$ on this node. Constraint $C_2$ indicates that at the arrival of the current $s_{i,j}$, the total memory resource allocation of all tasks hosted by the destination node cannot exceed the resource quota $mem$ on this node. Constraint $C_3$ states that the $deadline_{s_{i,j}}$ of a workflow task $s_{i,j}$ needs to be satisfied regardless of whether it is scheduled and executed by an edge node or a cloud node. Variables $p$ and $l$ need to satisfy $1\leq p \leq m$ and $1\leq l \leq n$.

\section{Cloud-edge Workflow Scheduling Engine}
This section elaborates on the cloud-edge workflow scheduling engine, including architecture design and core subsystem introduction. 

\subsection{Architecture Design}
\begin{figure*}[h] 
 \centering
 \includegraphics[width=1\textwidth]{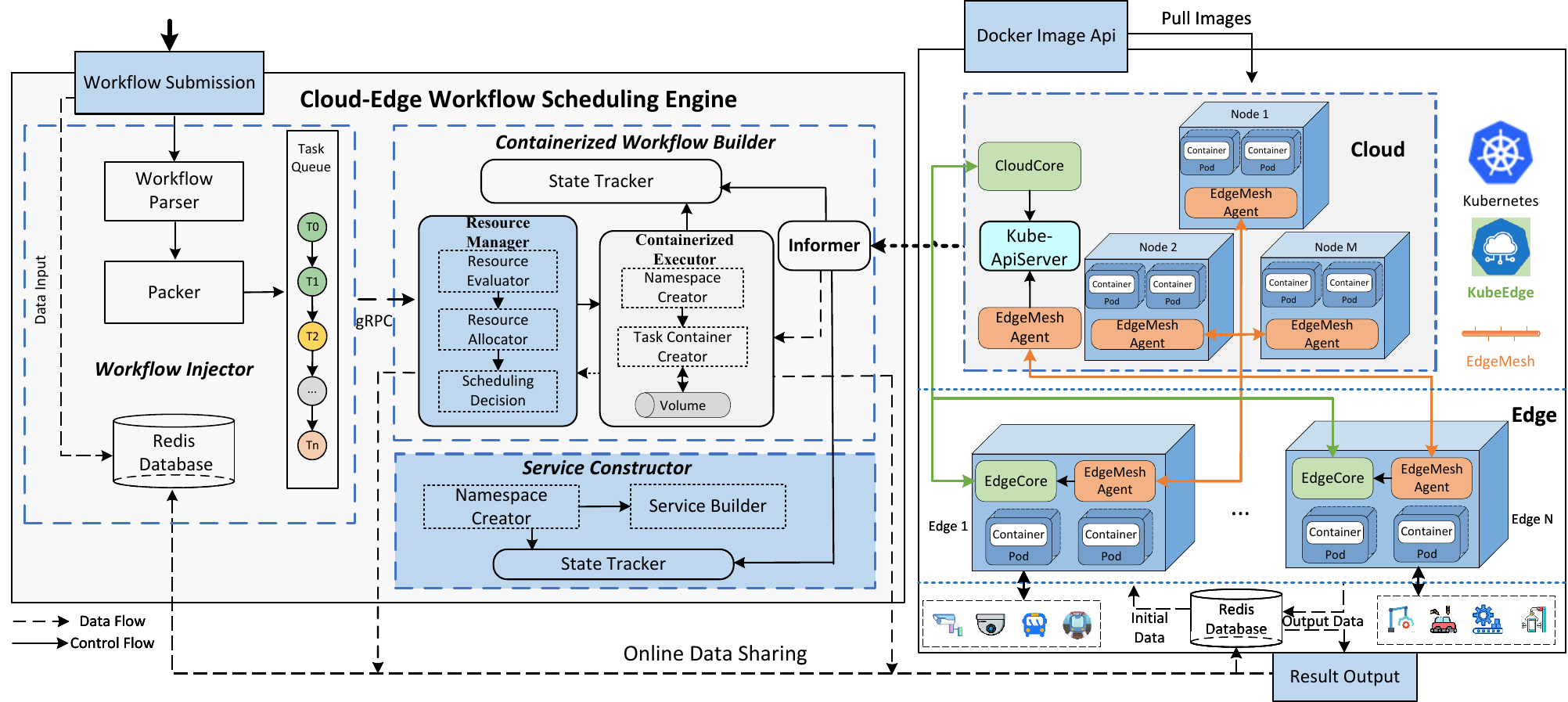}
 \caption{Cloud-edge workflow scheduling engine}
 \label{cloud_edge_wf_scheduling_engine}
\end{figure*}

The cloud-edge scenario builds on K8s and KubeEdge frameworks, incorporating EdgeMesh components. The left part of Fig.~\ref{cloud_edge_wf_scheduling_engine} shows the detailed architecture of the cloud-edge workflow scheduling engine, and the right part presents the cloud-edge-end three-layer architecture of the cloud-edge scenario.

The cloud-edge workflow scheduling engine realizes the management, scheduling, and containerized execution of workflow tasks across cloud-edge nodes. The engine provides a front-end interface for public or private clouds. The cloud-edge containerization scheduling of workflow tasks follows the priority dependency of tasks. Under the condition of satisfying the workflow SLAs, the resource evaluation algorithm~\ref{algo:resource-evaluation}, resource allocation algorithm~\ref{algo:resource-allocation}, and scheduling decision are used to ensure the maximum utilization of cloud-edge resources. It provides a one-click deployment capability and automated software operation process for the KubeEdge framework.

\subsection{Core Subsystems}
The cloud-edge workflow scheduling engine contains three core subsystems: Workflow Injector, Containerized Workflow Builder, and Service Constructor.

\subsubsection{Workflow Injector}
Users can define workflows and deliver them in the form of cloud services~\cite{shan2023kubeadaptor}. The workflow Injector completes the parsing and packaging of workflows and injects workflows into the Containerized Workflow Builder via gRPC communication. During workflow delivery, workflow data is populated into the Redis database.

\subsubsection{Containerized Workflow Builder}
\label{sec:cwb}
The Containerized Workflow Builder contains Resource Manager, Containerized Executor, State Tracker, and Informer.

\textbf{Resource Manager} contains Resource Evaluator, Resource Allocator, and Scheduling Decision. Once the workflow task arrives at the Containerized Workflow Builder, the Resource Evaluator evaluates whether the delay constraints of the current task can be satisfied at the cloud or edge node according to the resource evaluation algorithm~\ref{algo:resource-evaluation}. Herein, the calculation of task execution time on cloud or edge nodes evolves into the calculation of available resources, that is, whether there are sufficient resources available in a specific timeslot $p$ to determine whether the current node meets the delay constraint of this task. In order to make full use of the cloud-edge node resources, the Resource Allocator adopts the resource scaling strategy~\cite{shan2023adaptive} and allocates required resources of the current task through the resource allocation algorithm~\ref{algo:resource-allocation}. The Scheduling Decision subsystem is to present the created task Pod to the Kube-scheduler of K8s for scheduling.

{\textbf{Containerized Executor} creates the workflow namespace and the PVC under the current namespace, generates the task Pod of the corresponding workflow, and realizes the data sharing between tasks under the same namespace through the Volume and NFS mechanism. The data of the created task Pod is populated into the Redis database in time.

\textbf{State Tracker} utilizes the List-Watch mechanism to provide resource state information such as Namespace, Pod, and PVC to the Containerized Executor.

\textbf{Informer} is to synchronize resource objects and events between the K8s apiserver~\footnote[3]{The K8s apiserver serves as the front end and is a component of the K8s control plane that exposes the K8s API.} and the Informer local cache and provides the State Tracker with up-to-date resource state information.

\subsubsection{Service Constructor}
The Service Constructor consists of Namespace Creator, Service Builder, and State Tracker.
 
\textbf{Namespace Creator} is functionally the same as the Namespace Creator in the Containerized Workflow Builder and is responsible for creating the namespace of the injected workflow. Resource isolation space is provided for the Service creation of subsequent workflow tasks.

\textbf{Service Builder} is responsible for creating the corresponding Service for each task under the workflow namespace, exposing the Service in ClusterIp mode~\cite{k8sService2023}. Workflow task Pod realizes task communication between tasks in cloud-edge scenarios by means of Service ClusterIp.

\textbf{State Tracker} is similar to the function of the Status Tracker within the Containerized Workflow Builder and is responsible for monitoring the state information of the namespace and Service resource objects through the ListWatch mechanism.

\section{Algorithm Implementation}
This section proposes the cloud-edge resource hybrid management strategy, integrates the cloud and edge node resources, and designs the resource evaluation and resource allocation algorithms to solve the resource coordination issue in workflow containerized scheduling of cloud-edge scenarios. For cloud-edge containerized workflow scheduling, this section proposes the distributed functional role strategy and the computing power hierarchical division strategy to implement the task horizontal roaming and vertical offloading algorithms against the application coordination issue.

\subsection{Resource Collaboration}
The cloud-edge workflow scheduling engine manages the cloud-edge resources through the KubeEdge framework and schedules as many workflow tasks as possible to execute on the cloud or edge nodes, aiming to maximize the resource utilization of cloud-edge nodes. However, this does not mean that all workflow tasks will run. Regardless of the cloud or edge node, if this node accepts multiple workflow tasks at a timeslot, it will lead to insufficient resources for these tasks to meet the task deadline and be scheduled. Whether the task execution time on the cloud or edge nodes can satisfy the task delay constraint depends on the available resources. The resource quota of the assigned task depends on the Resource Manager of the workflow scheduling engine, and the resource allocation of the Resource Manager depends on the task number hosted by the current cloud or edge node.
We evaluate whether an arriving workflow task is acceptable by calculating whether the current node has enough available resources. In order to maximize the resource utilization of cloud or edge nodes, the resource scaling method proposed in our former work~\cite{shan2023adaptive} is adopted, and the resource request parameters of workflow tasks are appropriately adjusted to make cloud or edge nodes accommodate more workflow task without affecting current workflow task.

\subsubsection{Cloud-Edge Resource Hybrid Management Strategy}
The cloud-edge workflow scheduling engine is deployed on the cloud in a containerized way. In order to realize resource evaluation and cloud-edge workflow scheduling, it is necessary to manage cloud-edge resources in the KubeEdge cluster uniformly. KubeEdge can extend K8s' container orchestration capabilities to the edge and help K8s obtain the resource objects of the edge node through the resource API methods. Herein, a data dictionary $ResidualResourceMap$ stores the remaining resources of all nodes in the cloud-edge cluster and is defined in Fig.~\ref{cloud_edge_node_residual_resource}.
\begin{figure}[h] 
 \centering
 \includegraphics[width=0.3\textwidth]{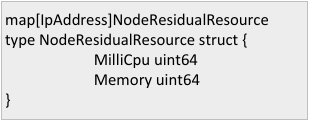}
 \caption{Data dictionary definition of the remaining resources at cloud or edge node}
 \label{cloud_edge_node_residual_resource}
\end{figure}

Herein, IpAddress is a string type and represents the IP address of the cloud-edge node. NodeResidualResource is a structural data structure that contains two $uint64$ type parameters of milliCpu and Memory representing the remaining CPU and memory resources on the current node, respectively. With the help of the API method of cloud-edge node resource operations, we can obtain PodLister and NodeLister. By calling the resource discovery algorithm~\cite{shan2023adaptive}, using PodLister, NodeLister, and ResidualMap, we are able to get ResidualResourceMap of cloud edge nodes. The ResidualResourceMap data dictionary provides the residual resource data of cloud-edge nodes for resource evaluation and achieves the purpose of hybrid resource management of cloud-edge nodes.

\subsubsection{Residual Resource Calculation For Cloud-edge Nodes}

The execution status of task Pod has Pending, Running, Succeeded, Failed, and Unknown~\cite{ilyushkin2017experimental}. According to the Pod characteristics of K8s, task Pods in Pending and Running states are considered to calculate the occupied resources. The set of task Pods in Pending and Running states on the current node is denoted as $S^p$ and $S^r$, respectively.

When the workflow task $s_{i,j}$ arrives at the cloud-edge workflow scheduling engine, the Resource Evaluator within the Resource Manager evaluates whether to accept this task according to the remaining resource amount of the current node and the total resource demand of the current task and the concurrent tasks in its lifecycle. Using the resource discovery algorithm, the CPU and memory resources occupied by the workflow task Pod active in the current node can be obtained as follows.
\begin{eqnarray}
  \label{eq_allocatable_cpu_mem_resources}
  \begin{split}
  &R_{used}^{cpu}=\sum\limits_{t \in S^{p}\cap S^{r}}Re^{cpu}(t),\\
  &R_{used}^{mem}=\sum\limits_{t \in S^{p}\cap S^{r}}Re^{mem}(t),
\end{split}
\end{eqnarray}

By using the resource discovery algorithm, the allocatable CPU and memory resources at the current node can be represented as $R^{cpu}_{allocatable}$ and $R^{mem}_{allocatable}$. The difference between the allocatable resources of the current node and the occupied resources is the remaining resources of this node.

\subsubsection{Resource Evaluation Algorithm}
The Resource Evaluator simulates workflow $w_i$ to further execute a lifecycle of the current task $s_{i,j}$ to predict the future concurrent tasks~\cite{ilyushkin2017experimental}. By searching the Redis database, we obtain the set $S$ of workflow tasks to be scheduled to the current node. Then, the set $S$ is searched for the workflow tasks (i.e., concurrent tasks) to be started within the lifetime of the current task $s_{i,j}$ (between the task start time $P^{s_{i,j}}_{start}$ and $P^{s_{i,j}}_{max}$). We denote this task set as $S^{compete}_{s_{i,j}}$. The total CPU and memory resource requirements of all workflow tasks in this set are calculated as follows.
\begin{eqnarray}
  \label{eq_alltasks_cpu_mem_requests}
  \begin{split}
  &Re_{compete}^{cpu}=\sum\limits_{t \in S_{s_{i,j}}^{compete}}Re^{cpu}(t),\\
  &Re_{compete}^{mem}=\sum\limits_{t \in S_{s_{i,j}}^{compete}}Re^{mem}(t).
\end{split}
\end{eqnarray}

The Resource Evaluator can determine whether the current node can accept the workflow task $s_{i,j}$ based on the following conditions:
\begin{eqnarray}
  \label{eq_cpu_mem_resource_decisions}
\begin{split}
  &R_{allocatable}^{cpu}\geq Re_{s_{i,j}}^{cpu}+R_{used}^{cpu}+Re_{compete}^{cpu},\\
  &R_{allocatable}^{mem}\geq Re_{s_{i,j}}^{mem}+R_{used}^{mem}+Re_{compete}^{mem}.
\end{split}
\end{eqnarray}

When the allocatable resources of the current node satisfy (\ref{eq_cpu_mem_resource_decisions}), the workflow scheduling engine schedules the task $s_{i,j}$ to the current node. 

\begin{algorithm}[!htbp]\caption{Resource Evaluation Algorithm}
\label{algo:resource-evaluation}
  \begin{algorithmic}[1]
    \STATE \textbf{Input}~~$s_{i,j}$;
    \STATE \textbf{Output}~~True or False;
    \STATE Initialize $R_{allocatable}^{cpu}$, $R_{allocatable}^{mem}$, $Re_{compete}^{cpu}$, $Re_{compete}^{mem}$, $R_{used}^{cpu}$ and $R_{used}^{mem}$ as zero;
    \STATE Use (\ref{eq_wf_task_resource_requirement}) to compute $Re_{s_{i,j}}^{cpu}$ and $Re_{s_{i,j}}^{mem}$ of task $s_{i,j}$;
    \STATE Populate $s_{i,j}$'s data into Redis and update the start time and deadline of task $s_{i,j}$;
    \STATE Check if the parent task of the current task $s_{i,j}$ is completed, otherwise, wait;
    \STATE Create the namespace and PVC of $w_i$ containing $s_{i,j}$;
    \STATE Obtain $R_{allocatable}^{cpu}$ and $R_{allocatable}^{mem}$ of the node to which $s_{i,j}$ will be scheduled;
    \STATE Obtain $R_{used}^{cpu}$ and $R_{used}^{mem}$ of the node to which $s_{i,j}$ will be scheduled;
    \FOR{task $t$ to be scheduled on the current node $\in$ $S$}
       \STATE // $t$ hasn't been executed yet
       \IF{$t.AliveStatus$ == false}
         \STATE // These tasks with equal labels belong to the same node
         \IF{$t.Labels$ == $s_{i,j}.Labels$}
           \IF{$t.start$~$\in$~$\left[P_{start}^{s_{i,j}},P_{max}^{s_{i,j}} \right]$}
             \STATE $R_{compete}^{cpu}$+=$Re^{cpu}(t)$;
             \STATE $R_{compete}^{mem}$+=$Re^{mem}(t)$;
           \ENDIF
         \ENDIF
       \ENDIF
    \ENDFOR
    \IF{$R_{allocatable}^{cpu}$$\geq$ $Re_{s_{i,j}}^{cpu}$+$R_{used}^{cpu}$+$Re_{compete}^{cpu}$~and~\\~~~~$R_{allocatable}^{mem}$$\geq$ $Re_{s_{i,j}}^{mem}$+$R_{used}^{mem}$+$Re_{compete}^{mem}$}
      \STATE return True;
    \ELSE
      \STATE return False;
    \ENDIF
  \end{algorithmic}
\end{algorithm}

Algorithm~\ref{algo:resource-evaluation} is the resource evaluation algorithm, which aims to provide the decision of scheduling workflow tasks to cloud or edge nodes. In line 4, the resource requirement of the current task $s_{i,j}$ is calculated using (\ref{eq_wf_task_resource_requirement}). When the workflow task arrives, line 5 populates the workflow task $s_{i,j}$'s data into the Redis database and updates the start time and deadline of the task. Then, line 6 checks to see if the current task's successor is complete. Once the conditions are available for execution, line 7 creates the namespace and PVC for this task. Lines 8 and 9 calculate $R_{allocatable}^{cpu}$, $R_{allocatable}^{mem}$, $R_{used}^{cpu}$ and $R_{used}^{mem}$ of the node to which $s_{i,j}$ will be scheduled, respectively.
Lines 10 to 21 traverse workflow task set $S$ in the Redis to obtain the total resource requirements $R_{compete}^{cpu}$ and $R_{compete}^{mem}$ of these tasks to be executed during the lifetime of task $s_{i,j}$. In line 22, True is returned when the available resources of the node to which task $s_{i,j}$ will be scheduled to satisfy (\ref{eq_cpu_mem_resource_decisions}). The time complexity of this algorithm is $O\big(\max \{ \left|V\right|\times\left|Pod.List\right|, \left|S\right| \} \big )$.

\subsubsection{Resource Allocation Algorithm}
\label{sec:resource-allocation}
This section maximizes node resource utilization by making the cluster nodes host more tasks, which continues the case that the resource evaluation algorithm~\ref{algo:resource-evaluation} returns False. When (\ref{eq_cpu_mem_resource_decisions}) cannot be satisfied at the same time or both, in order to make cloud or edge nodes accommodate more workflow tasks and maximize the resource utilization of cloud or edge nodes, we calculate the resource demand according to the following equations: 
\begin{eqnarray}
  \label{eq-resource-scaling-rate}
  \begin{split}
&Re_{cut}^{cpu,s_{i,j}} = Re_{s_{i,j}}^{cpu} \cdot \Bigg ( \frac{R_{allocatable}^{cpu}-R_{used}^{cpu}}{Re_{s_{i,j}}^{cpu}+Re_{compete}^{cpu}}\Bigg),\\
&Re_{cut}^{mem,s_{i,j}} = Re_{s_{i,j}}^{mem} \cdot \Bigg ( \frac{R_{allocatable}^{mem}-R_{used}^{cpu}}{Re_{s_{i,j}}^{mem}+Re_{compete}^{mem}}\Bigg ).
  \end{split} 
\end{eqnarray}

The CPU and memory resource requirements scaled by (\ref{eq-resource-scaling-rate}) should meet the minimum resource requirements of the current task $s_{i,j}$ to run. The minimum resource requirements of task $s_{i,j}$ are expressed as $Re_{s_{i,j}}^{cpu,min}$ and $Re_{s_{i,j}}^{mem,min}$, which depend on resource amount operated by the resource load tool Stress within the Pod of task $s_{i,j}$.
Within the task Pod, Stress allocates 1 CPU fork and $Re_{s_{i,j}}^{mem,min}$ memory as resource load. In order to ensure the smooth running of task Pod, the minimum memory resource amount of the task Pod corresponding to task $s_{i,j}$ needs to be increased by a constant $\beta$=20 after experimental testing (the minimum CPU resource amount is not considered, because CPU is a compressible resource). The resource amount $Re_{s_{i,j}}^{mem,min}+\beta$ ensures that the Pod corresponding to task $s_{i,j}$ will not fail to execute because the Stress tool runs out of memory resources.

\begin{algorithm}[!htbp]\caption{Resource Allocation Algorithm}
\label{algo:resource-allocation}
  \begin{algorithmic}[1]
    \STATE \textbf{Input}~~$s_{i,j}$;
    \STATE \textbf{Output}~~$R_{s_{i,j},cpu}^{allocated}$ and $R_{s_{i,j},mem}^{allocated}$;
      \STATE Obtain $R_{allocatable}^{cpu}$, $R_{allocatable}^{mem}$, $R_{used}^{cpu}$, $R_{used}^{mem}$, $Re_{compete}^{cpu}$ and $Re_{compete}^{mem}$ by using algorithm~\ref{algo:resource-evaluation};
      \STATE Obtain $Re_{cut}^{cpu,s_{i,j}}$ and $Re_{cut}^{mem,s_{i,j}}$ through (\ref{eq-resource-scaling-rate});
      \STATE Define $R_{allocatable}^{cpu}$$\geq$ $Re_{s_{i,j}}^{cpu}$+$R_{used}^{cpu}$+$Re_{compete}^{cpu}$ as $A_1$, $R_{allocatable}^{mem}$$\geq$ $Re_{s_{i,j}}^{mem}$+$R_{used}^{mem}$+$Re_{compete}^{mem}$ as $A_2$;
      \STATE Define $\neg$ as logical negation and $\wedge$ as logical and;

      \IF{$A_1$ $\wedge$ $A_2$}
        \STATE $R_{s_{i,j},cpu}^{allocated}$ = $Re_{s_{i,j}}^{cpu}$;~~$R_{s_{i,j},mem}^{allocated}$ = $Re_{s_{i,j}}^{mem}$;
      \ELSE
        \IF{$\neg$ $A_1$ $\wedge$ $A_2$}
            \STATE $R_{s_{i,j},cpu}^{allocated}$ = $Re_{cut}^{cpu,s_{i,j}}$;~~$R_{s_{i,j},mem}^{allocated}$ = $Re_{s_{i,j}}^{mem}$;
        \ELSE
          \IF{$A_1$ $\wedge$ $\neg$ $A_2$}
            \STATE $R_{s_{i,j},cpu}^{allocated}$ = $Re_{s_{i,j}}^{cpu}$;~~$R_{s_{i,j},mem}^{allocated}$ = $Re_{cut}^{mem,s_{i,j}}$;
          \ELSE
            \STATE $R_{s_{i,j},cpu}^{allocated}$ = $Re_{cut}^{cpu,s_{i,j}}$;~~$R_{s_{i,j},mem}^{allocated}$ = $Re_{cut}^{mem,s_{i,j}}$;
          \ENDIF
        \ENDIF
      \ENDIF

      \IF{$R_{s_{i,j},cpu}^{allocated}$ $\geq$ $Re_{s_{i,j}}^{cpu,min}$~and~$R_{s_{i,j},mem}^{allocated}$ $\geq$ 
    $\left( Re_{s_{i,j}}^{mem,min}+\beta \right)$}
        \STATE return $R_{s_{i,j},cpu}^{allocated}$ and $R_{s_{i,j},mem}^{allocated}$;
      \ELSE
        \STATE Call the horizontal roaming algorithm \ref{algo:task-roamming-algorithm} to search for other edge nodes with sufficient resources in the edge scene where task $s_{i,j}$ is located;
      \ENDIF

  \end{algorithmic}
\end{algorithm}

Algorithm~\ref{algo:resource-allocation} is the resource allocation algorithm for edge or cloud nodes in cloud-edge scenarios. This algorithm is able to obtain the allocatable resource $R_{s_{i,j},cpu}^{allocated}$ and $R_{s_{i,j},mem}^{allocated}$ on the node to be scheduled for task $s_{i,j}$.
Lines 3 to 5 obtain the various parameters and conditions of this algorithm. When the total resource demand of task $s_{i,j}$ and its concurrent tasks in the future lifecycle is not greater than the residual resources of this node, lines 7 to 8 allocate resources according to the resource demand of task $s_{i,j}$.
When the total CPU resource demand of task $s_{i,j}$ and its concurrent tasks in the future lifecycle is greater than the residual CPU resource of this node, lines 10 to 11 allocate the scaled CPU resource $Re_{cut}^{cpu,s_{i,j}}$ and memory resource $Re_{s_{i,j}}^{mem}$ to task $s_{i,j}$. When the total memory resource demand of task $s_{i,j}$ and its concurrent tasks in the future lifecycle is greater than the residual memory resource of this node, the scaled memory resource $Re_{cut}^{mem,s_{i,j}}$ and $Re_{s_{i,j}}^{cpu}$ are allocated to task $s_{i,j}$ in line 13 and line 14. 
When neither condition A1 nor A2 holds, lines 15 and 16 present the scaled CPU and memory resources. If the allocated resource $R_{s_{i,j},cpu}^{allocated}$ and $R_{s_{i,j},mem}^{allocated}$ do not meet the comparison conditions in line 20 and the horizontal roaming algorithm~\ref{algo:task-roamming-algorithm} is invoked to search for other edge nodes with sufficient resources in edge scenario of task $s_{i,j}$. Then, the resource allocation of task $s_{i,j}$ is performed on the edge node obtained by the search. Resource allocation algorithm~\ref{algo:resource-allocation}'s time complexity depends on the resource revaluation algorithm~\ref{algo:resource-evaluation}'s time complexity $O\big(\max \{ \left|V\right|\times\left|Pod.List\right|, \left|S\right| \} \big )$.

\subsubsection{Scheduling Decision}
When the task $s_{i,j}$ obtains $R_{s_{i,j},cpu}^{allocated}$ and $R_{s_{i,j},mem}^{allocated}$ allocated by the resource allocation algorithm~\ref{algo:resource-allocation}, the Containerized Executor creates the namespace of the workflow in which the task located, and the corresponding Pod for the task $s_{i,j}$. Finally, the default scheduler of K8s, i.e., the Kube-scheduler, schedules the task Pod to the corresponding cloud or edge node.

\subsection{Application Collaboration}
This section uses the edge-edge horizontal roaming and cloud-edge vertical offloading scheme of workflow task containers to realize the workflow containerization scheduling in cloud-edge scenarios and to deal with the difficulty of cloud-edge application collaborative management.

\subsubsection{Task Horizontal Roaming Scheme}
In the data collection scenario of IoT, the sensor devices in the perception layer collect a large amount of data. Due to the delay sensitivity and data privacy of some data, there is a large amount of communication delay in batch uploading to the cloud, which cannot meet the real-time requirements of applications. 
Therefore, it is necessary to complete data processing at the edge nodes near the device to save data transmission time. A data processing model is formed for edge nodes to deploy data-intensive applications and cloud nodes to deploy diagnosis and decision-making platforms.

\textbf{Distributed function role policy.}
According to different functional roles, edge nodes host data-intensive applications, and cloud nodes run diagnosis and decision-making platforms. 
When faced with a large volume of data-collecting scenarios, new business loads will continuously gather on edge nodes, and data collecting and task processing will exhaust the resources of edge nodes, resulting in insufficient resource allocation for subsequent tasks. 
When the task load hosted by the node is heavy, resource shortage will cause part of the workflow task to fail (task Pod \verb|OOMKilled|), so the task horizontal roaming algorithm is needed to migrate the current failed task Pod to the same edge scene. Other nodes of the same role in this scene are used to host this task and ensure the smooth operation of the workflow. The Kube-scheduler uses a label-matching mechanism to schedule workflow task Pod for the cloud-edge scenario. The task label has a data dictionary structure, represented by $map\left[Key: IpAddress\right]$. $Key$ represents the label keyword of the scheduling node and belongs to a node-set $\left\{node\mbox{-}1,node\mbox{-}2,...,node\mbox{-}m, edge\mbox{-}1,edge\mbox{-}2,...,edge\mbox{-}n\right\}$. Each task Pod keeps the same label as its host node, which ensures that the Kube-scheduler schedules task Pods to its matching node.

\textbf{Task horizontal roaming algorithm.}
Algorithm~\ref{algo:task-roamming-algorithm} implements the horizontal roaming of the task Pod with \verb|OOMKilled| state, returning the labels of the destination node in the same edge scenario. 
\begin{algorithm}[!htbp]\caption{Task Horizontal Roaming Algorithm}
  \label{algo:task-roamming-algorithm}
  \begin{algorithmic}[1]
    \STATE \textbf{Input}~~$s_{i,j}$;
    \STATE \textbf{Output}~~Label of the destination node;
    \STATE The Informer within workflow scheduling engine automatically captures the $OOMKilled$ task $s_{i,j}$ and executes the delete operation;
    \STATE Obtain the clusterNodeLabelMap(map[key][]IpAddress) of all node labels in cloud-edge cluster;
    \STATE Obtain the \verb|labelKey| and \verb|IPAddress| of task $s_{i,j}$;
    \STATE Obtain the remaining resource list of the cluster nodes;
    \FOR {$\forall$ IPAddress $\in$ clusterNodeLabelMap[labelKey]}
        \IF{IPAddress $\neq$ $s_{i,j}$.IPAddress}
          \STATE Search for the same role node with the largest residual source in the same edge scenario from the node remaining resource list where the task $s_{i,j}$ is located;
          \STATE Update the label key of task $s_{i,j}$ as the iPAddress of the edge node with the largest residual resources;       
        \ENDIF
    \ENDFOR
    \STATE Update the label of task $s_{i,j}$ in Redis database;
    \STATE Return the Labels of the node to which task $s {i,j}$ will roam; 
  \end{algorithmic}
\end{algorithm}

Line 3 describes that the workflow scheduling engine captures the task Pod of the \verb|OOMKilled| state in real-time through the Informer module and removes this task. Line 4 obtains the clusterNodeLabelMap of the labels of all nodes in the cluster. Line 5 gets the keywords and key values of the task labels for \verb|OOMKilled| tasks. Then, line 6 obtains the remaining resource list of the cluster nodes. Line 7 traverses the $clusterNodeLabelMap[labelKey]$ of the node label in the same edge scenario. Line 8 to line 10 exclude the current node to search the node with the largest residual resources and record the node IP address of this node, which is the horizontal roaming node of task $s_{i,j}$. In line 13, the label of task $s_{i,j}$ in the Redis database is updated, and the label of this roaming node is returned. Time complexity of algorithm~\ref{algo:task-roamming-algorithm} is $O\big( \left|V\right|\times\left|Pod.List\right| \big )$.

\subsubsection{Task Vertical Offloading Scheme}
In the industrial IoT field, in order to maintain workshop equipment, it is necessary to deploy remote equipment monitoring tasks on edge nodes near the devices, continuously obtain equipment status data collected by sensors for preprocessing, and send the results to the diagnosis platform. The diagnosis platform processes data preprocessing results from edge node monitoring tasks. 
The diagnosis platform task can drift between the cloud node and the edge node in accordance with the specific monitoring requirements of the manufacturer. 

\textbf{Computing power hierarchical division strategy.}
In intensive device-monitoring mode, product testing and deployment are the main requirements. In order to reduce the data transmission delay between monitoring tasks and diagnosis platform tasks, diagnosis platform tasks can be offloaded to edge nodes with low computing power levels. In the regular device-monitoring mode, the daily business application of the device is the primary requirement. To save the resources of the edge nodes, diagnosis platform tasks can be offloaded to the cloud nodes with high computing power.

The task vertical offloading algorithm follows the idea of computing power hierarchical division, which can reasonably offload tasks running on cloud or edge nodes and give full play to the advantages of cloud-edge resources. To test the task vertical offloading scheme, the cloud-edge workflow scheduling engine appropriately scales the resource requirements of tasks according to the resource scaling strategy, and the vertical offloading process of \verb|OOMKilled| task Pod caused by insufficient resources is to simulate the offloading of diagnostic platform tasks from edge nodes to cloud nodes. 
Due to resource heterogeneity of cloud-edge nodes, the data dictionary of cluster task image is defined as $map\left[Ip: ImageAddress\right]$, where $Ip$ represents the IP address set of cluster nodes. $ImageAddress$ stands for the image address of the diagnostic platform task on the corresponding node. Since the task vertical offloading scheme realizes the task offloading from the edge node to the cloud node, which involves the different processor architectures of the nodes, the task offloading should consider the change of the image address for the diagnosis platform task.

\textbf{Task vertical offloading algorithm.}
Algorithm~\ref{algo:task-offloading-algorithm} implements the vertical offloading of the task Pod with \verb|OOMKilled| state and returns the label of the destination cloud node corresponding to the edge scene.

\begin{algorithm}[!htbp]\caption{Task Vertical Offloading Algorithm}
  \label{algo:task-offloading-algorithm}
  \begin{algorithmic}[1]
    \STATE \textbf{Input}~~$s_{i,j}$;
    \STATE \textbf{Output}~~The label of offloading node;
    \STATE The Informer within workflow scheduling engine automatically captures the $OOMKilled$ task $s_{i,j}$ and executes the delete operation;
    \STATE Obtain the clusterNodeLabelMap(map[key][]IpAddress) of all node labels in cloud-edge cluster;
   \STATE Obtain the NodeTaskImageAddressMap(map[Ip][]Image\\Address) of task images for all nodes in cloud-edge cluster;
   \STATE Obtain the \verb|labelKey| and \verb|IPAddress| of task $s_{i,j}$;
    \STATE Obtain the remaining resource list of the cluster nodes;
    \FOR {$\forall$ labelKey $\in$ clusterNodeLabelMap}
        \STATE Get clusterNodeLabelMap of cloud node matching labelKey;
        \FOR{$\forall$ IPAddress $\in$ clusterNodeLabelMap[labelKey]}
            \IF{IPAddress $\neq$ $s_{i,j}$.IPAddress}
              \STATE Search for the cloud node with the largest remaining resource in the cloud scenario that corresponds to the edge scenario of task $s_{i,j}$;
              \STATE Record the IPAddress of cloud node with maximum residual resources as  the label key value of task $s_{i,j}$ and update the task image address NodeTaskImageAddressMap[IPAddress].   
            \ENDIF
        \ENDFOR
    \ENDFOR
    \STATE Update new labels of task $s_{i,j}$ in Redis database;
    \STATE Return the label of offloading node of task $s_{i,j}$;
  \end{algorithmic}
\end{algorithm}

In line 3 of algorithm~\ref{algo:task-offloading-algorithm}, the cloud-edge workflow scheduling engine captures the \verb|OOMKilled| task Pod in real-time through the Informer module and removes this task. Line 4 gets the clusterNodeLabelMap (map[key][]IpAddress) of the labels of all nodes in the cluster, and line 5 obtains the NodeTaskImageAddressMap (map[Ip][]ImageAddress) for all cluster nodes. Then, line 7 obtains the residual resource list of the cluster nodes. Line 8 traverses the clusterNodeLabelMap of the cluster nodes, and for each labelKey, line 9 obtains the clusterNodeLabelMap[labelKey] of the cloud node that matches the labelKey. Then line 10 traverses clusterNodeLabelMap[labelKey], and lines 11 to 13 search the cloud node with the largest residual resources (corresponding to the edge scenario of the node where task $s_{i,j}$ is located) excluding the current node, and records the IP address of the cloud node. This cloud node is the vertical offloading node of task $s_{i,j}$. In line 17, the label of task $s_{i,j}$ in Redis data is updated, and finally, the label of this offloading node is returned. Since len(clusterNodeLabelMap) in line 8 is less than $V$ (the number of cluster nodes). The clusterNodeLabelMap[labelKey] in line 9 of the algorithm is the set of IP addresses of nodes in a certain edge scenario, and this set must also be smaller than len(Pod.List) in the cluster. So the time complexity of algorithm~\ref{algo:task-offloading-algorithm} is $O\big( \left|V\right|\times\left|Pod.List\right| \big )$.

\section{Experimental Evaluation and Analysis}
In this section, we use the cloud-edge workflow scheduling engine to run the customized IoT workflow to test the performance of the related algorithms. 
We sequentially present the experimental setup and performance evaluation.

\subsection{Experimental Setup}
This section details the experimental setup, including the experiment testbed, workflow instance, workflow arrival pattern, evaluation metrics, and baseline.

\subsubsection{Experimental Testbed}
As shown in Fig.~\ref{cloud_edge_test_bed}, the cloud consists of K8s clusters, and the edge includes four Raspberry PI edge nodes. We use KubeEdge to build the cloud-edge cluster, which integrates the cloud, edge, and device end containing sensors and other collecting devices. The edges and device end belong to distinct edge scenarios, such as edge-1 and edge-2. The cloud-edge workflow scheduling engine is deployed in the K8s cluster in the cloud, with continuous injection of workflows. For IoT workflow application scenarios, delay-sensitive tasks and tasks with data privacy need to be scheduled to the edge nodes near the device by the cloud-edge workflow scheduling engine for execution. Table~\ref{tab:testbed_Configuration} describes the software and hardware configuration of the cloud-edge testbed. 
\begin{figure}[h]
 \centering
 \includegraphics[width=3.5in]{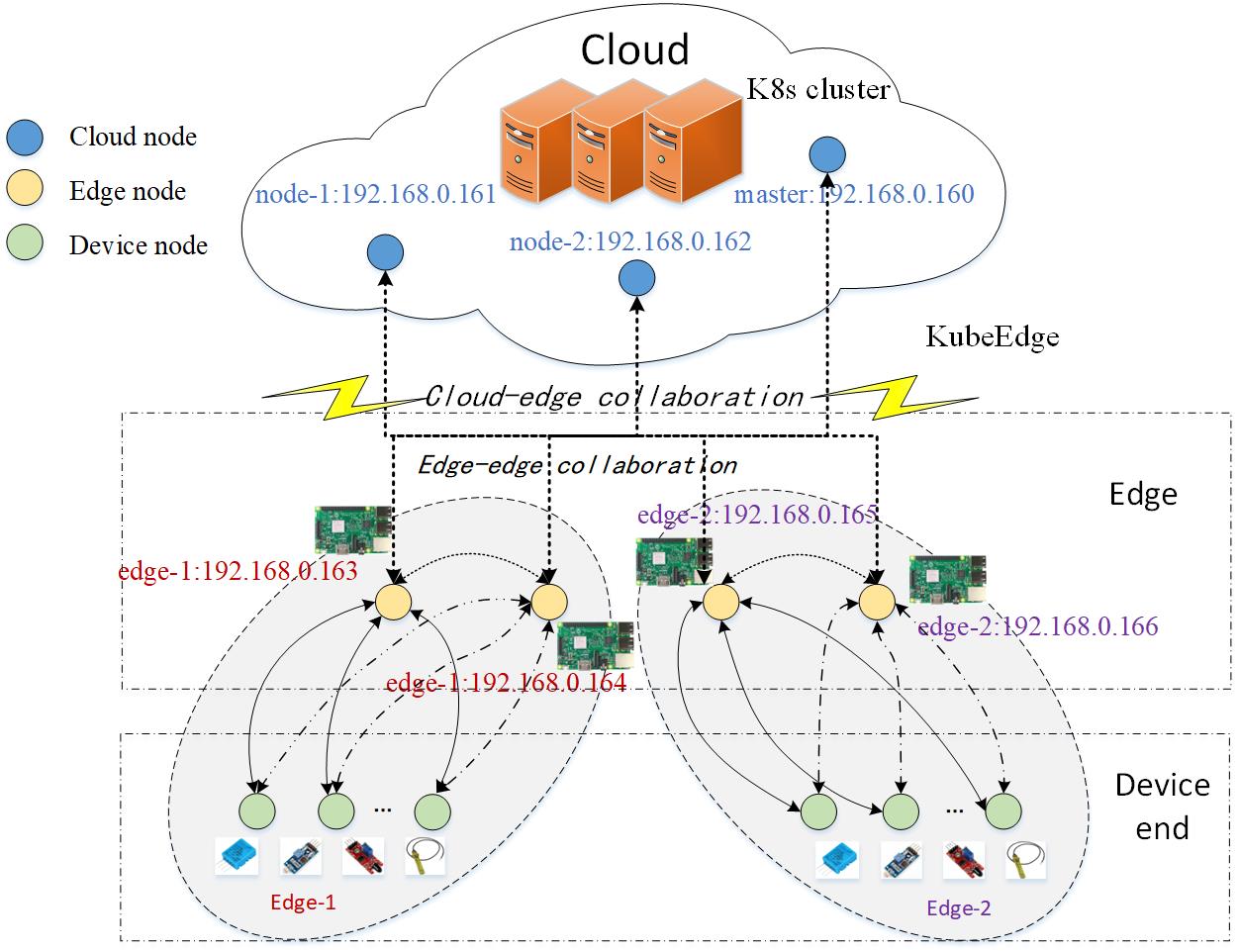}
 \caption{Cloud-edge testbed.}
 \label{cloud_edge_test_bed}
\end{figure}


\begin{table*}[!t]
 \footnotesize
	\centering
	\caption{Configuration of cloud-edge experimental testbed}
	\renewcommand{\multirowsetup}{\centering}
	\begin{tabular}
    {|m{1.3cm}<{\centering}|m{2.8cm}<{\centering}|m{3cm}<{\centering}|m{2.3cm}<{\centering}|m{1cm}<{\centering}|m{2cm}<{\centering}|}
    \hline
		\textbf{Node Category}&\textbf{Node Number}&\textbf{CPU Specifications}&\textbf{Software}&\multicolumn{2}{c|}{\textbf{Hardware Configuration}} \\
		\hline
		\multirow{4}{*}{Cloud Node}&{master:192.168.0.160}&{}&{ubuntu~20.4}&{CPU}&{1 Core} \\
   &{node-1:192.168.0.161}&{Intel(R) Core(TM)}&{K8s 1.19.6}&{Memory}&{2 GB} \\
		&{node-2:192.168.0.162}&{i7-1165G7@2.80GHz}&{KubeEdge 1.9.1}&{Bandwidth}&{100Mbps}\\
    &{}&&{Docker18.9.6}&{NIC}&{Intel PRO/1000}\\
    \hline
		\multirow{3}{*}{Edge Node}&{edge-1:192.168.0.163}&{Raspberry Pi4 }&{ubuntu~20.4}&{CPU}&{4 Cores} \\
    &{edge-1:192.168.0.164}&{Model B}&{K8s 1.19.6}&{Memory}&{2 GB} \\
		&{edge-2:192.168.0.165}&{Broadcom BCM2835}&{KubeEdge 1.9.1}&{Bandwidth}&{100 Mbps}\\
		&{edge-2:192.168.0.166}&{Cortex-A72@1.5GHz}&{Docker18.9.6}&{NIC}&{2.4GHz WiFi}\\
    \hline
	\end{tabular}
	\label{tab:testbed_Configuration}
\end{table*}


\subsubsection{Workflow Instance}
In this section, we design a universal workflow suitable for IoT application scenarios. The workflow topology includes multiple branch combinations of the DAG graph (in-tree, out-tree, fork-join, etc.), which aims to simulate the relationship between businesses in IoT applications as much as possible.
\begin{figure}[h] 
 \centering
 \includegraphics[width=2.5in]{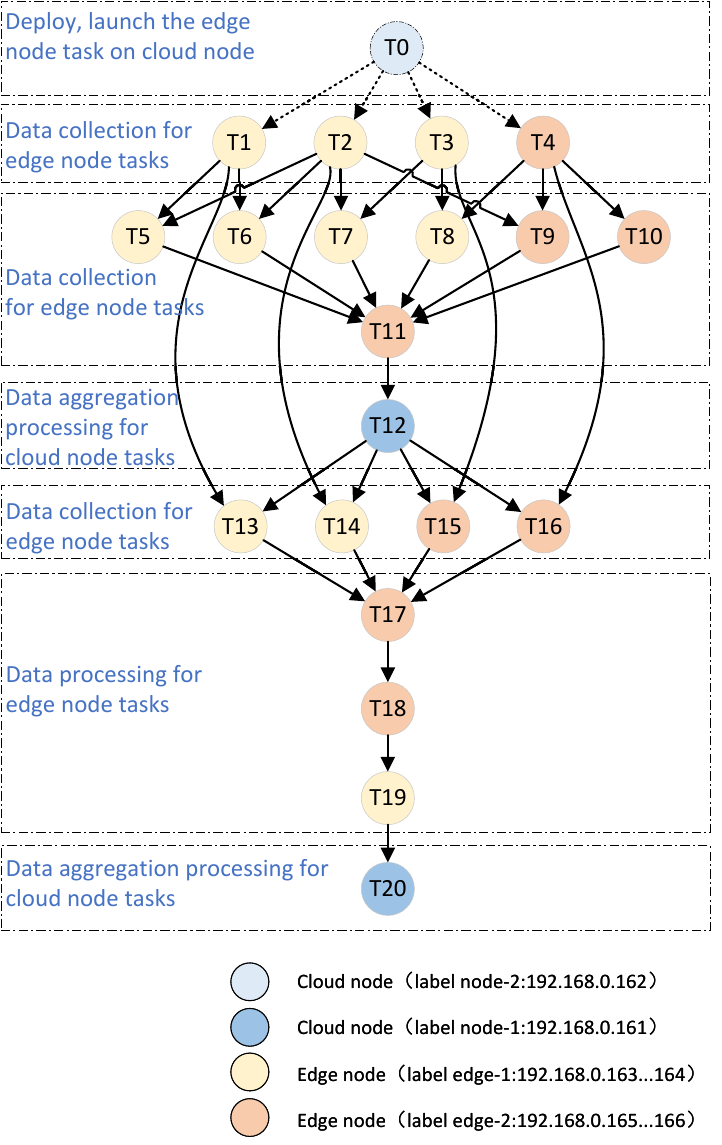}
 \caption{Workflow instance.}
 \label{workflow_instance}
\end{figure}

Fig.~\ref{workflow_instance} depicts a universal workflow topology in an IoT scenario. The workflow task $T_0$ located at the top layer is responsible for workflow deployment, the start of the edge node task, and runs on the cloud node. Then, the data collection tasks of the second layer edge nodes (edge nodes in edge-1 and edge-2) are triggered following workflow dependencies and converge to the third layer edge nodes (edge nodes in edge-1 and edge-2) for data processing. Next, the data flows into task $T_{12}$ of the fourth layer cloud node, which is responsible for further data processing. 
$T_{12}$ continues to trigger the edge node tasks at the fifth layer to collect data, and the data is aggregated to the edge node at the sixth layer to complete processing in turn. Finally, the data flows into the task $T_{20}$ of the cloud node to provide data processing results for users to make decisions. 
For the data collection, data processing, and decision-making tasks in the IoT application workflow~(Fig.~\ref{workflow_instance}), we simulate the execution process of these tasks through the Stress tool. By injecting the relevant parameters into the task container, the Stress tool will execute the corresponding resource load, simulating the running of the real task. 

\subsubsection{Workflow Arrival Pattern}
We use the following three workflow arrival patterns to simulate the intensity of workflow arrival. The task horizontal roaming and vertical offloading algorithm tests use constant arrival patterns to inject workflows. The Workflow Injector module sends a specified number of workflows to the Containerized Workflow Builder every 300 seconds. For instance, a total of 10 workflows, each batch number $y=2$, sending once every 300 seconds, a total of 5 times.

\textbf{Constant arrival pattern.} The number of workflow requests arrives in a constant manner. The Workflow Injector module injects two workflow requests to the Containerized Workflow Builder every 300 seconds~(that is $y=2$). The whole process continuously injects five times, a total of 10 workflows, as shown in Fig.~\ref{fig:resource-usage-curve}(a).

\textbf{Linear arrival pattern.} The number of workflow requests is injected into the Containerized Workflow Builder in an increasing linear function~($y=k*x+d$, where $y$ is the amount of injected workflow requests and $d$ takes the initial value of 1). The injected workflow requests are increased by $k$~($k=1$) every 300 seconds, and the whole process sends injections four times, a total of 10 workflows, as shown in Fig.~\ref{fig:resource-usage-curve}(b).

\textbf{Pyramid arrival pattern.} The curve of the number of workflow requests conforms to a pyramid shape. The experiment starts with a small number of concurrent workflow requests (1 workflow) until it grows to a randomly chosen larger value (limited to cloud-edge resources, here taken as 3), as shown in Fig.~\ref{fig:resource-usage-curve}(c). The number of injected workflow requests grows by one every 300 seconds until a peak. Once reaching the peak, the number immediately falls back to its initial value in the same way, and the process executes repeatedly until the total number of workflow requests~(i.e., 17).

\subsubsection{Evaluation Metrics}
The evaluation metrics of cloud-edge workflow scheduling optimization include total workflow duration, average workflow duration, CPU resource usage, and memory resource usage. The evaluation metrics for task horizontal roaming and vertical offloading algorithms contain the lifecycle of horizontal roaming or vertical offloading tasks, task number, and resource allocation during task roaming and offloading.

\textbf{Total workflow duration (min).} It represents the total duration of all injected workflows, that is, the time taken by workflow injection from the first workflow request to the completion of the last workflow execution.

\textbf{Average workflow duration (min).} This metric represents the average execution time of a single workflow, that is, the time taken by each workflow from the beginning of the first task to the end of the last task.

\textbf{Resource usage.} It contains the average resource usage of CPU and memory. This metric reflects the average resource usage over the total duration of all injected workflows in the KubeEdge cluster. 

\textbf{Lifecycle of horizontal roaming or vertical offloading tasks.} This metric represents the time consumption of a \verb|OOMKilled| task from being captured to the end of roaming or offloading.

\textbf{Number of horizontal roaming or vertical offloading tasks.} This metric reflects the number of roaming and offloading tasks that occur during the injection of a certain number of workflows.

\textbf{Resource allocation in horizontal roaming or vertical offloading tasks.} This metric shows the execution effect of task horizontal roaming or vertical offloading algorithms and the fault-tolerant management ability of cloud-edge workflow scheduling engine through the resource allocation of roaming or offloading tasks.

\subsubsection{Baseline}
We use the resource allocation strategy in~\cite{shan2023kubeadaptor} as the baseline that does not consider the potential concurrent task requests in the lifecycle of the current task. It means that the resource allocation strategy in the baseline follows the First-Come-First-Serve (FCFS) scheme, and this scheme depends on the adequacy of the remaining resources on a node in the cloud-edge cluster. 
If the residual resource amount of the node is sufficient, the resource allocation strategy is implemented. Otherwise, it needs to wait for other task Pods on the node to complete and release resources to meet the resource requirements requested by the current task.

\subsection{Performance Evaluation}
This section depicts the execution process of the cloud-edge workflow scheduling engine running the IoT application workflow in the cloud-edge scenario, evaluates the performance of the KCES scheme, and verifies the effect of the task horizontal roaming and vertical offloading algorithms.

\subsubsection{Cloud-edge Workflow Scheduling Performance Analysis}
To verify the performance of KCES, we adopt the cloud-edge configuration in Fig.~\ref{cloud_edge_test_bed} and use the cloud-edge workflow scheduling engine to run the IoT workflow application (Fig.~\ref{workflow_instance}) on the KubeEdge cluster. The resource allocation algorithm in the KCES aims to make each cloud or edge node host as many workflow tasks as possible and maximize the resource utilization of cloud-edge nodes without violating the task deadline. When the allocated resources of the node where the current task resided do not meet the minimum amount of memory resource $Re_{s_{i,j}}^{mem,min}+\beta$, the task horizontal roaming algorithm~\ref{algo:task-roamming-algorithm} is triggered to search for other nodes with sufficient resources in the same scenario to host the current task. It alleviates the shortage of allocatable resources of the current node and also realizes load balancing under the same scenario. 
We evaluate the performance of KCES using total workflow duration, average workflow duration, CPU resource usage, and memory resource usage.
 \begin{table*}
  \footnotesize
 	\caption{Performance metric data of cloud-edge workflow scheduling}
  \label{table:cloud-edge-evaluation-performance}
   \centering
 \renewcommand\arraystretch{1} 
\begin{tabular}{|m{1cm}<{\centering}|m{4cm}<{\centering}|m{1.4cm}<{\centering}|m{1.4cm}<{\centering}|m{1.4cm}<{\centering}|m{1.4cm}<{\centering}|m{1.4cm}<{\centering}|m{1.4cm}<{\centering}|}
    \hline
 	\multirow{4}{*}{\bfseries{Workflow}}&\multirow{2}{*}{\bfseries{Metrics}}& \multicolumn{2}{c|}{\bfseries{Constant Arrival}} & \multicolumn{2}{c|}{\bfseries{Linear Arrival}} & \multicolumn{2}{c|}{\bfseries{Pyramid Arrival}} \\ 
 	\cline{3-8}
    &  & KCES & Baseline & KCES & Baseline & KCES  & Baseline  \\ 
   	\cline{2-8}
    & Number of workflow requests & \multicolumn{2}{c|}{10} & \multicolumn{2}{c|}{10} & \multicolumn{2}{c|}{17} \\  
   	\cline{2-8}
 \bfseries{Types} & Interval between two requests bursts/s & \multicolumn{2}{c|}{300} & \multicolumn{2}{c|}{300} & \multicolumn{2}{c|}{300} \\ 
  \hline 
 \multirow{10}{*}{\rotatebox{90}{\bfseries{IoT application workflow}}} &	\multirow{2}{*}{Total workflow duration (min)} & \multirow{2}{*}{26.3} & \multirow{2}{*}{30.5} & \multirow{2}{*}{25} & \multirow{2}{*}{33.1} & \multirow{2}{*}{45.8} & \multirow{2}{*}{49.8} \\ [0.7ex]

  & (Standard deviation) & ($\delta = 0.6$) & ($\delta = 1.1$) & ($\delta = 1.1$) & ($\delta = 2.0$) & ($\delta = 0.25$) & ($\delta = 0.94$) \\ [0.7ex]

  \cline{2-8}
  & \multirow{2}{*}{Average workflow duration (min)}  & \multirow{2}{*}{6.37} & \multirow{2}{*}{8.24} & \multirow{2}{*}{6.9} & \multirow{2}{*}{12.44} & \multirow{2}{*}{8.1} & \multirow{2}{*}{11.52} \\ [0.7ex]
  
  & (Standard deviation) & ($\delta = 0.38$) & ($\delta = 0.9$) & ($\delta = 1.03$) & ($\delta = 1.1$) & ($\delta = 0.56$) & ($\delta = 0.47$) \\ [0.7ex]
  
  \cline{2-8}
  &	\multirow{2}{*}{CPU resource usage}  & \multirow{2}{*}{0.06} & \multirow{2}{*}{0.05} & \multirow{2}{*}{0.06} & \multirow{2}{*}{0.05} & \multirow{2}{*}{0.06} & \multirow{2}{*}{0.05} \\ [0.7ex]
   
  & (Standard deviation) & ($\delta = 0.00$) & ($\delta = 0.00$) & ($\delta = 0.00$) & ($\delta = 0.00$) & ($\delta = 0.00$) & ($\delta = 0.00$) \\ [0.7ex]
  
  \cline{2-8}
  & \multirow{2}{*}{Memory resource usage}  & \multirow{2}{*}{0.16} & \multirow{2}{*}{0.15} & \multirow{2}{*}{0.16} & \multirow{2}{*}{0.15} & \multirow{2}{*}{0.17} & \multirow{2}{*}{0.15} \\ [0.7ex]
 	
  & (Standard deviation) & ($\delta = 0.01$) & ($\delta = 0.00$) & ($\delta = 0.01$) & ($\delta = 0.00$) & ($\delta = 0.00$) & ($\delta = 0.00$) \\ [0.7ex]
\hline
\end{tabular}
 \end{table*}

As shown in Table~\ref{table:cloud-edge-evaluation-performance}, "KCES" represents our proposed KCES in this paper, and "Baseline" stands for the baseline algorithm. Three arrival patterns inject 10, 10, and 17 workflows, respectively. Under different workflow arrival patterns, the KCES outperforms the baseline in total workflow duration, average workflow duration, and CPU and memory resource usage metrics. Table~\ref{table:cloud-edge-evaluation-performance} shows the mean and standard deviation $\delta$ of performance index data. The low standard deviation of various metrics in the table indicates that the dispersion of different metric data is low, which reflects the stability and reliability of the KCES.
The table shows that the CPU and resource usage data have little difference for each workflow arrival pattern, and the two resource usage curves are similar for each workflow arrival pattern, which is attributed to the same resource scaling ratio of CPU and memory.

Fig.~\ref{fig:resource-usage-curve} depicts the CPU and memory resource usage rate under three distinct arrival patterns for IoT application workflows. The resource curve used in each workflow arrival pattern in the figure generally ends later than the workflow request curve because each workflow has a deadline in the future, and some workflows are still waiting in the queue for execution until all injected workflows have been executed.

\begin{figure*}
	\centering
    \vspace{-0.5cm}
	\subfloat[Constant Arrival Pattern]{
		\begin{minipage}[t]{0.33\linewidth}
			\centering
			\includegraphics[width=2.4in]{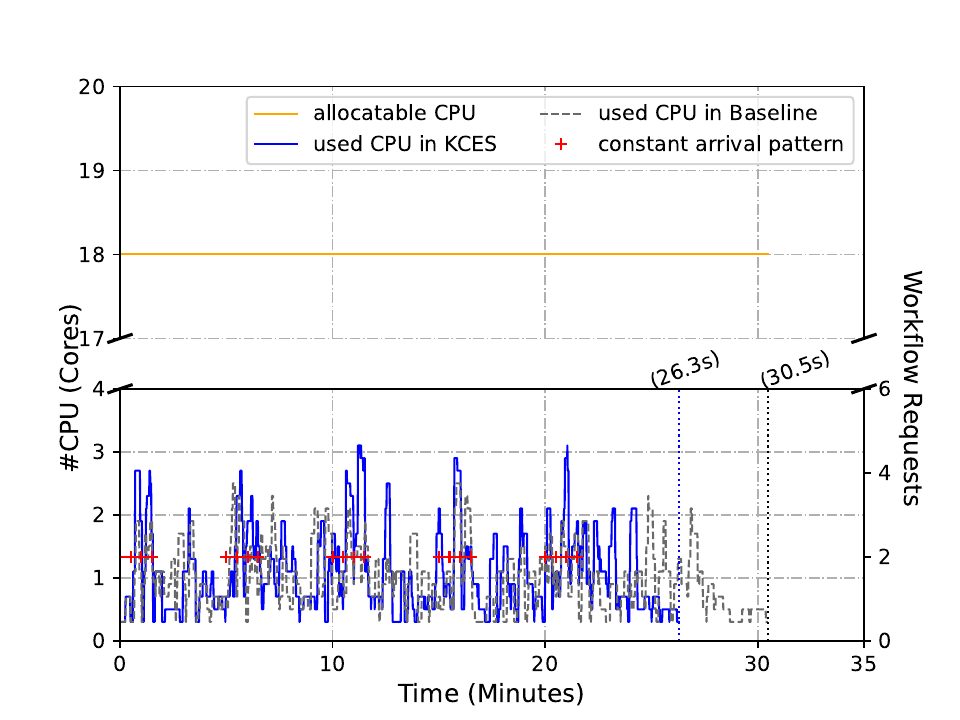}\\
			\vspace{0.02cm}
			\includegraphics[width=2.4in]{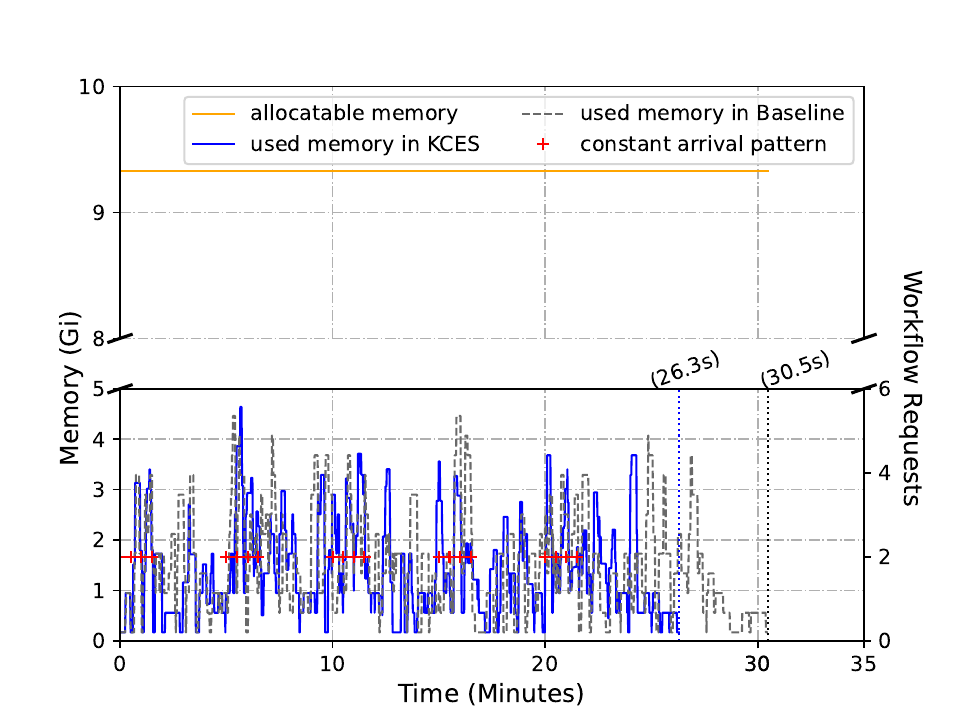}\\
			\vspace{0.02cm}
		\end{minipage}%
	}%
	\subfloat[Linear Arrival Pattern]{
		\begin{minipage}[t]{0.33\linewidth}
			\centering
			\includegraphics[width=2.4in]{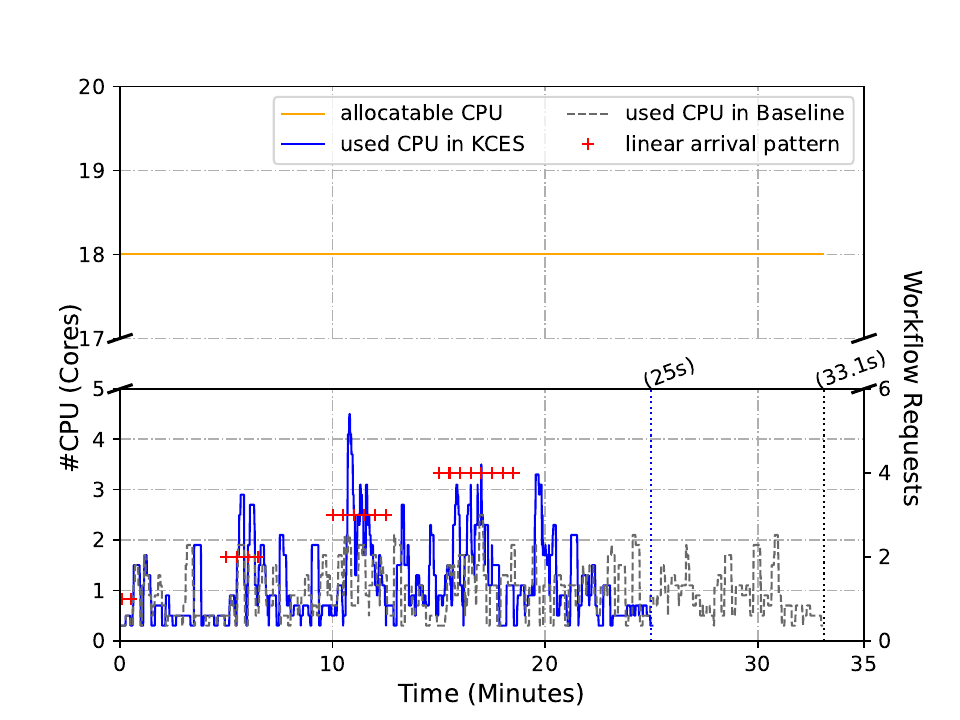}\\
			\vspace{0.02cm}
			\includegraphics[width=2.4in]{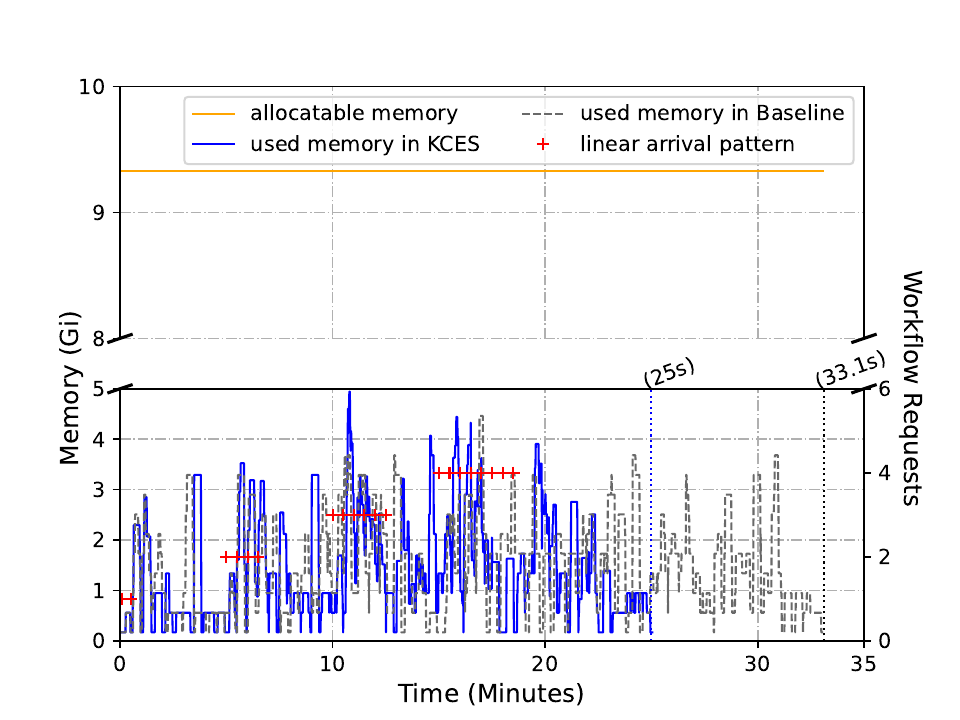}\\
			\vspace{0.02cm}
		\end{minipage}%
	}%
	\subfloat[Pyramid Arrival Pattern]{
		\begin{minipage}[t]{0.33\linewidth}
			\centering
			\includegraphics[width=2.4in]{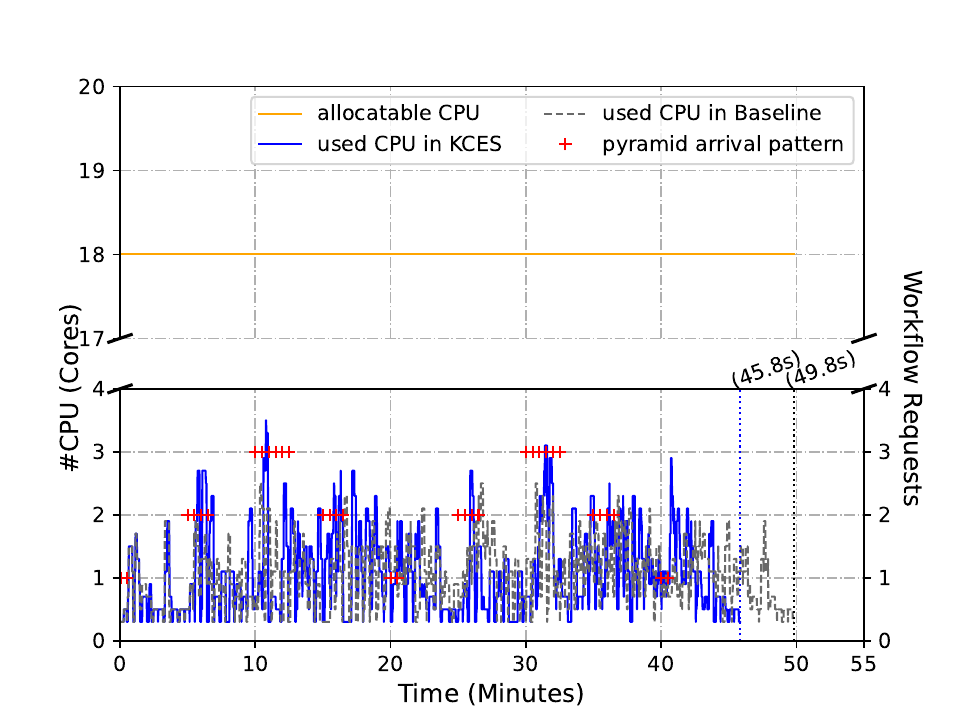}\\
			\vspace{0.02cm}
			\includegraphics[width=2.4in]{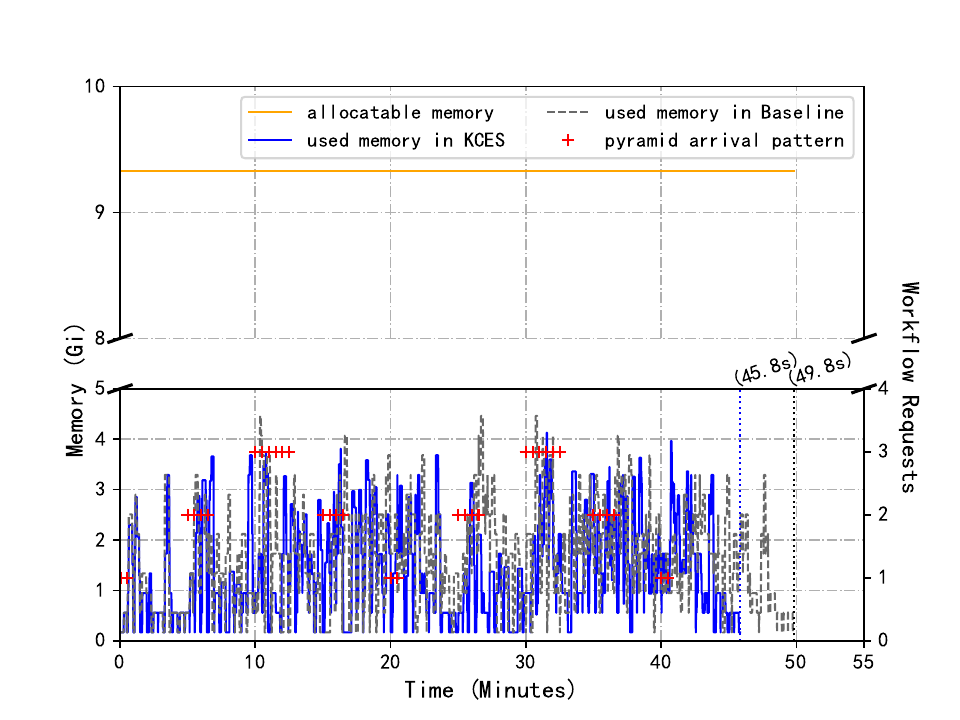}\\
			\vspace{0.02cm}
		\end{minipage}%
	}%
	\centering
	\caption{The CPU and memory resource usage rate under three distinct arrival patterns for Iot application workflows.}
  \vspace{-0.2cm}
	\label{fig:resource-usage-curve}
\end{figure*}

In the experimental setup, the IoT application workflow consists of 21 tasks (refer to Fig.~\ref{workflow_instance}), and workflow tasks will run on cloud or edge nodes. Compared to the baseline, for total workflow durations in Table~\ref{table:cloud-edge-evaluation-performance}, the KCES saves 13.8\% in constant arrival mode, 24.5\% in linear arrival mode, and 8\% in pyramid arrival mode. The KCES respectively saves 22.7\%, 44.5\%, and 29.7\% in average workflow duration under three different workflow arrival patterns compared to the baseline, as shown in Fig.~\ref{fig:resource-usage-curve}(a), (b) and (c), respectively. In linear arrival mode, the KCES obtains better performance in terms of total workflow duration and average workflow duration. It is due to that the concurrent degree of injected workflow requests is directly related to total workflow duration and average workflow duration. The higher the concurrency of injected workflow requests, the more workflow tasks run per timeslot, and the shorter the total workflow duration and average duration of a single workflow. For CPU and memory resource usage in Fig.~\ref{fig:resource-usage-curve}, KCES outperforms the baseline for each arrival pattern. Due to the limited resources in cloud-edge nodes, the injected workflow load is small, and the CPU and memory resource utilization under two strategies is not high. Compared to the baseline, KCES increases CPU resource usage by 20\% and memory resource usage by 6.6\% to 13.3\%.

From the resource usage curves (CPU and memory) of three workflow arrival patterns in Fig.~\ref{fig:resource-usage-curve}, we can observe that the resource usage peak of our KCES is higher than that of the baseline algorithm most of the time. Moreover, the peak of the resource usage curve is generally consistent with the centralized arrival of workflow requests. 
It is because the former uses a resource scaling method to adjust the resource requirements of potential future task requests during the lifecycle of the current task. Under the premise of ensuring the smooth operation of task Pods, the strategy starts task Pods on the current node as much as possible to speed up workflow execution efficiency. In addition, for the case of insufficient resource allocation of nodes, the task horizontal roaming algorithm alleviates the lack of resource allocation on the current node, realizes load balancing in the same scenario, and speeds up the execution efficiency of the workflow by searching other nodes with the same role with sufficient resources in the same scenario to host the task.
The baseline algorithm depends on the adequacy of the remaining resources on the cluster nodes. In the scenario of high workflow concurrency, the insufficient remaining resources of the nodes will cause the baseline algorithm to suffer from endless waiting and a lot of time waste, which prolongs the total workflow duration and average duration of a single workflow.

\subsubsection{Task Horizontal Roaming Performance Analysis}
To verify the horizontal roaming of workflow tasks in an edge scenario, the experiment in this section follows the cloud-edge configuration in Fig.~\ref{cloud_edge_test_bed}, and the cloud-edge workflow scheduling engine runs on cloud nodes. We evaluate the task horizontal roaming algorithm from the lifecycle of horizontal roaming tasks, the number of horizontal roaming tasks, and resource allocation.

\textbf{Lifecycle of horizontal roaming tasks.} 
The lifecycle of a horizontal roaming task refers to the total time consumption of the whole process of task creation, task \verb|OOMKilled| caused by insufficient resources, task deletion, task roaming, task recreation, task execution completion, and task deletion again. 
\begin{figure*}[htbp] 
 \centering
 \includegraphics[width=0.8\textwidth]{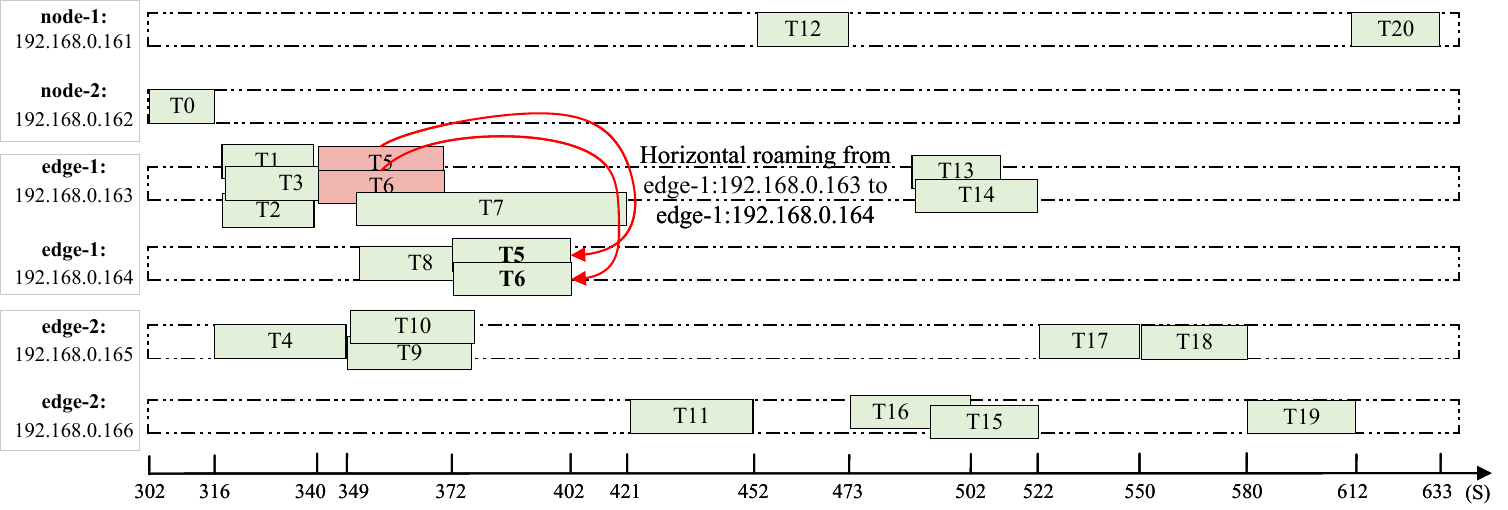}
 \caption{Task horizontal roaming process in workflow-3.}
 \label{task-horizontal-roaming-instance}
\end{figure*}
The experiment adopts the constant arrival pattern to inject four workflows in two batches every 300 seconds. This section selects a group of experimental data from multiple experiments to analyze the task roaming lifecycle metric. 

The experimental data show that two tasks of workflow-3 have occurred horizontal roaming ahead of the completion of four workflows.
As shown in Fig.~\ref{task-horizontal-roaming-instance}, the workflow-3's lifecycle is from the 302nd to the 633rd second of the experiment process, consuming 331 seconds. The task roaming lifecycle of $T5$ and $T6$ lasts from the 340th second to the 402nd second for 62 seconds. T5 and T6 migrate horizontally from $192.168.0.163$ in the edge scenario edge-1 to $192.168.0.164$ in this scenario. In addition, Fig.~\ref{task-horizontal-roaming-instance} shows that the task roaming process of $T5$ and $T6$ in workflow-3 does not affect the smooth operation of the whole workflow in the cloud-edge scenario.

\textbf{Number of horizontal roaming tasks.}
In the experiment, the Workflow Injector module injected $b$ workflows concurrently every 300 seconds in constant arrival mode, and $N$ workflows needed to inject $N/b$ batches in total. Under the effect of the task horizontal roaming algorithm, the cloud-edge workflow scheduling engine executed five groups of experimental tests $(N=2,b=1)$, $(N=4,b=2)$, $(N=6,b=3)$, $(N=8,b=4)$ and $(N=10,b=5)$. The total number of injected workflow tasks is 42, 84, 126, 168, and 210, respectively, and the successful execution rate of workflow tasks in each group of experiments is $100\%$.
\begin{figure}[htbp] 
 \centering
 \includegraphics[width=3.3in]{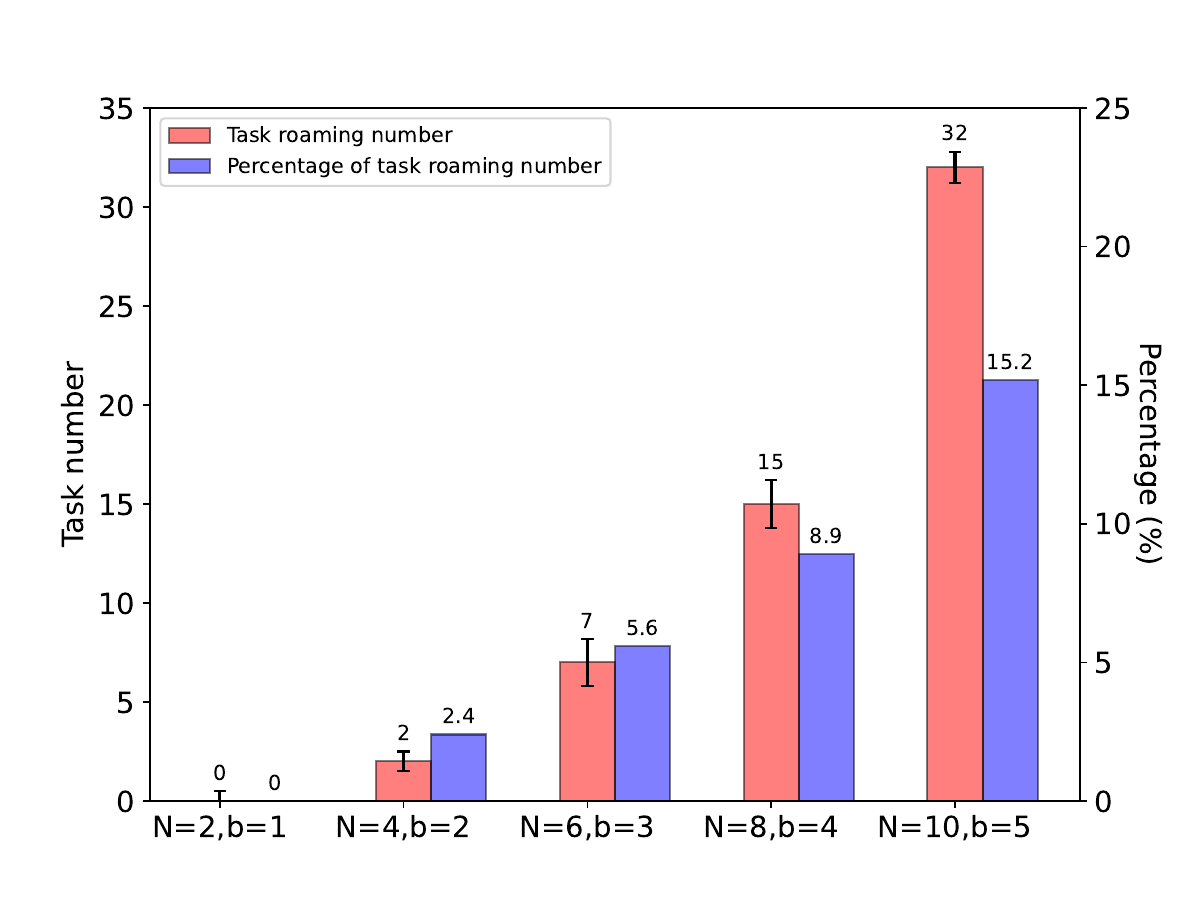}
 \caption{Relationship between workflow scale and the number of horizontal roaming tasks}
 \label{task_roaming_number}
\end{figure}

Fig.~\ref{task_roaming_number} depicts the relationship between workflow scale and the number of horizontal roaming tasks for five group experiments. In Fig.~\ref{task_roaming_number}, the error bar~(standard deviation) of roaming task number changes slightly, which reflects that the Resource Allocator module of the cloud-edge workflow scheduling engine has high stability when facing a high workflow workload.
We can observe that the $(N=2,b=1)$ group experiment does not have task horizontal roaming due to the small workflow load, sufficient resource allocation, and no \verb|OOMKilled| task Pod. With the growth of workflow scale, the load in the cloud-edge cluster increases, and the resources become tighter, leading to the gradual increase of the task roaming number and the percentage of task roaming.

In the experiment data group $(N=10,b=5)$, the number of horizontal roaming tasks has been as high as 32, and the percentage of horizontal roaming tasks reaches $15.2\%$. Under the effect of the resource scaling strategy in Resource Allocator, each cloud-edge node can accommodate more tasks, and some task Pods become \verb|OOMKilled| state because of resource shortage.
With the help of the task horizontal roaming algorithm, the cloud-edge workflow scheduling engine can automatically capture the \verb|OOMKilled| task Pod and transfer the load to other nodes in the same scenario to realize the resource reallocation of the \verb|OOMKilled| task Pod and ensure the normal execution of the task Pod.

\textbf{Resource allocation for horizontal roaming tasks.} 
We choose task $T9$ of workflow-0 with horizontal roaming event in experiment data group $(N=10, b=5)$ to analyze resource allocation. At the beginning of the task lifecycle, the cloud-edge workflow scheduling engine simulates the current task $T9$ running a lifecycle, predicts the possible concurrent tasks in the future, and uses the resource scaling strategy to host as many task Pods as possible, aiming to maximize the resource utilization of cloud-edge nodes.
\begin{figure}[htbp] 
 \centering
 \includegraphics[width=3.3in]{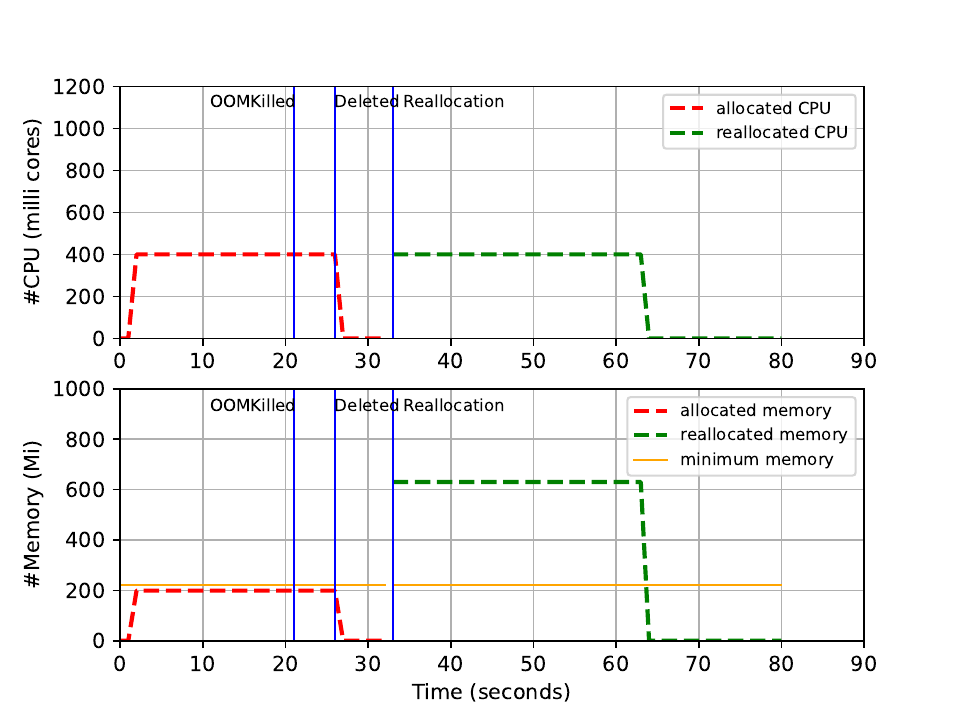}
 \caption{Resource reallocation for horizontal roaming task.}
 \label{task_roaming_task_oom}
\end{figure}

Currently, the allocated resources for task $T9$ are 400 millicores CPU and 199 Mi memory, since the memory request required by the internal Stress tool of $T9$ in the workflow definition is 200Mi ($Re_{s_{i,j}}^{mem,min}$), the memory request for normal execution of task9's Pod is not less than $Re_{s_{i,j}}^{mem,min}+\beta$, i.e., 220Mi ($\beta=20$).
As shown in Fig.~\ref{task_roaming_task_oom}, at the 21st second, task $T9$ encounters \verb|OOMKilled| due to insufficient memory resource allocation. Next, the cloud-edge workflow scheduling engine captures this task and calls the task horizontal roaming algorithm~\ref{algo:task-roamming-algorithm} to remove this task while obtaining other edge nodes suitable for hosting this task.
At the 26th second, the Pod corresponding to task $T9$ is removed from the cluster. Then, the workflow scheduling engine invokes the resource allocation algorithm at the 33rd second to restart the resource allocation for $T9$. The reallocated resources $(CPU:400m,Memory:630Mi)$ are enough to ensure the smooth operation of $T9$, and finally, at the 63rd second, the workflow scheduling engine deletes the successfully executed task $T9$.

\subsubsection{Task Vertical Offloading Performance Analysis}
Using the experimental testbed in Fig.~\ref{cloud_edge_test_bed} and the cloud-edge workflow scheduling engine, this section evaluates the task vertical offloading algorithm from the lifecycle of vertical offloading tasks, the number of vertical offloading tasks, and resource allocation.

\textbf{Lifecycle of vertical offloading tasks.} 
The lifecycle of a vertical offloading task refers to the total time consumption of the whole process of task creation, task \verb|OOMKilled| caused by insufficient resources, task deletion, task offloading, task recreation, task execution completion, and task deletion again. 
The experiment verifies the task vertical offloading algorithm of the KCES and adopts the constant arrival pattern to inject four workflows in two batches every 300 seconds. This section selected a group of data from multiple experiments to analyze the task offloading lifecycle metric.
\begin{figure*}[h] 
 \centering
 \includegraphics[width=0.8\textwidth]{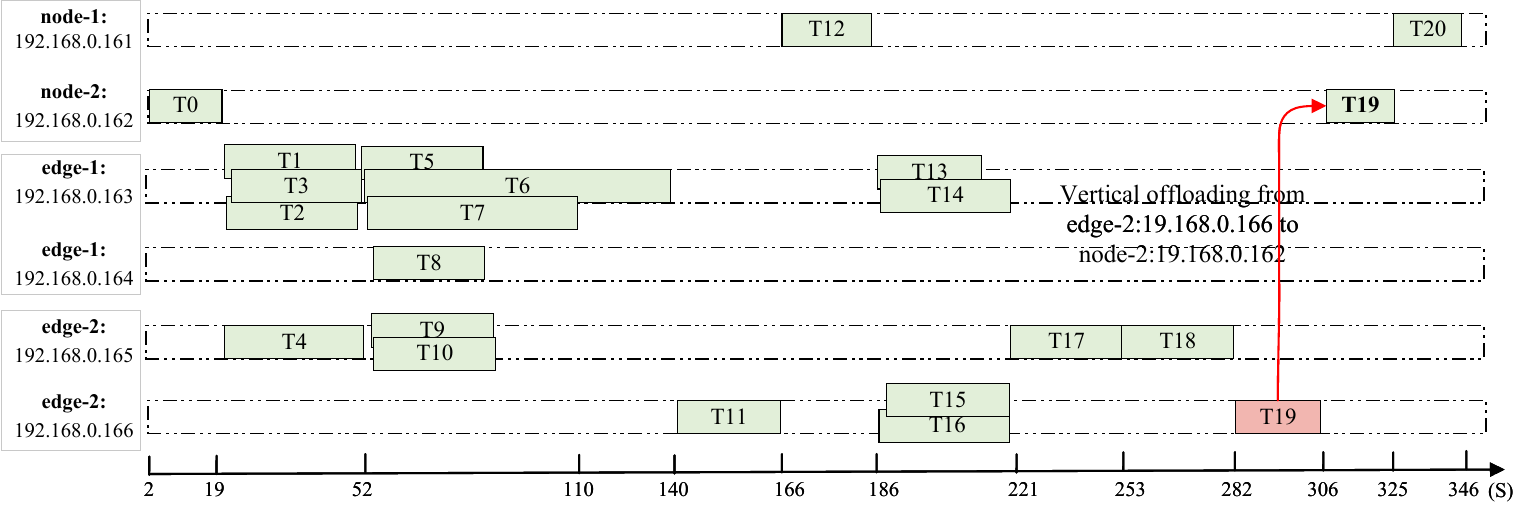}
 \caption{Task vertical offloading process in workflow-0.}
 \label{fig_task_offloading_wf0_1task}
\end{figure*}
\begin{figure*}[h] 
 \centering
 \includegraphics[width=0.8\textwidth]{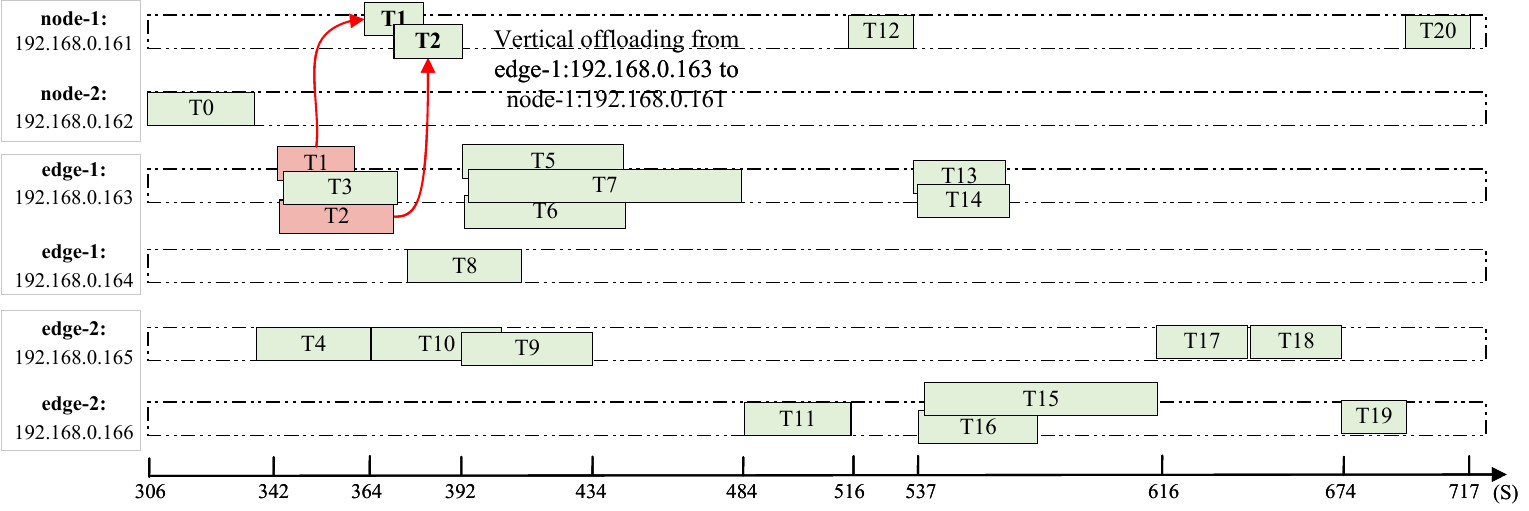}
 \caption{Task vertical offloading process in workflow-2.}
 \label{fig_task_offloading_wf2_2tasks}
\end{figure*}

As shown in Fig.~\ref{fig_task_offloading_wf0_1task} and Fig.~\ref{fig_task_offloading_wf2_2tasks}, task $T19$ of workflow-0 and task $T1$ and $T2$ of workflow-2 have been vertically offloaded as of the completion of four workflows. Fig.~\ref{fig_task_offloading_wf0_1task} shows that the lifecycle of workflow-0 takes 344 seconds from the 2nd second to the 346th second. The task $T19$ of workflow-0 undergoes vertical offloading from the 282nd second to the 325th second, offloading from 192.168.0.166 of edge scenario edge-2 to cloud node node2:192.168.0.162, and the lifecycle takes 43 seconds.
Fig.~\ref{fig_task_offloading_wf2_2tasks} shows that the lifecycle of workflow-2 takes 411 seconds from the 306th second to the 717th second, and the vertical offloading process of its task $T1$ lasts from the 342nd second to the 380th second, and the lifecycle takes 38 seconds. The vertical offloading process of task $T2$ lasts from the 342nd second to the 392nd second, and the lifecycle takes 50 seconds.
Due to insufficient resource allocation, task $T1$ and task $T2$ encounter \verb|OOMKilled|, the task vertical offloading algorithm offloads two tasks from 192.168.0.163 in edge scenario edge-1 to cloud node node1:192.168.0.161. In addition, Fig.~\ref{fig_task_offloading_wf0_1task} and Fig.~\ref{fig_task_offloading_wf2_2tasks} also show that the task vertical offloading process of workflow-0 and workflow-2 does not affect the successful execution of the workflow in the cloud-edge scenario.

\textbf{Number of vertical offloading tasks.}
Similar to the task horizontal roaming experiment, the task vertical offloading experiment also used a constant arrival pattern to inject $b$ workflows concurrently every 300 seconds by the Workflow Injection module, and a total of $N/b$ batches were injected. Five sets of $(N=2,b=1)$, $(N=4,b=2)$, $(N=6,b=3)$, $(N=8,b=4)$ and $(N=10,b=5)$ experimental tests were performed. The total number of injected tasks is 42, 84, 126, 168, and 210, respectively, and the successful execution rate of each group of experimental workflow tasks reaches $100\%$.
\begin{figure}[htbp] 
 \centering
 \includegraphics[width=3.3in]{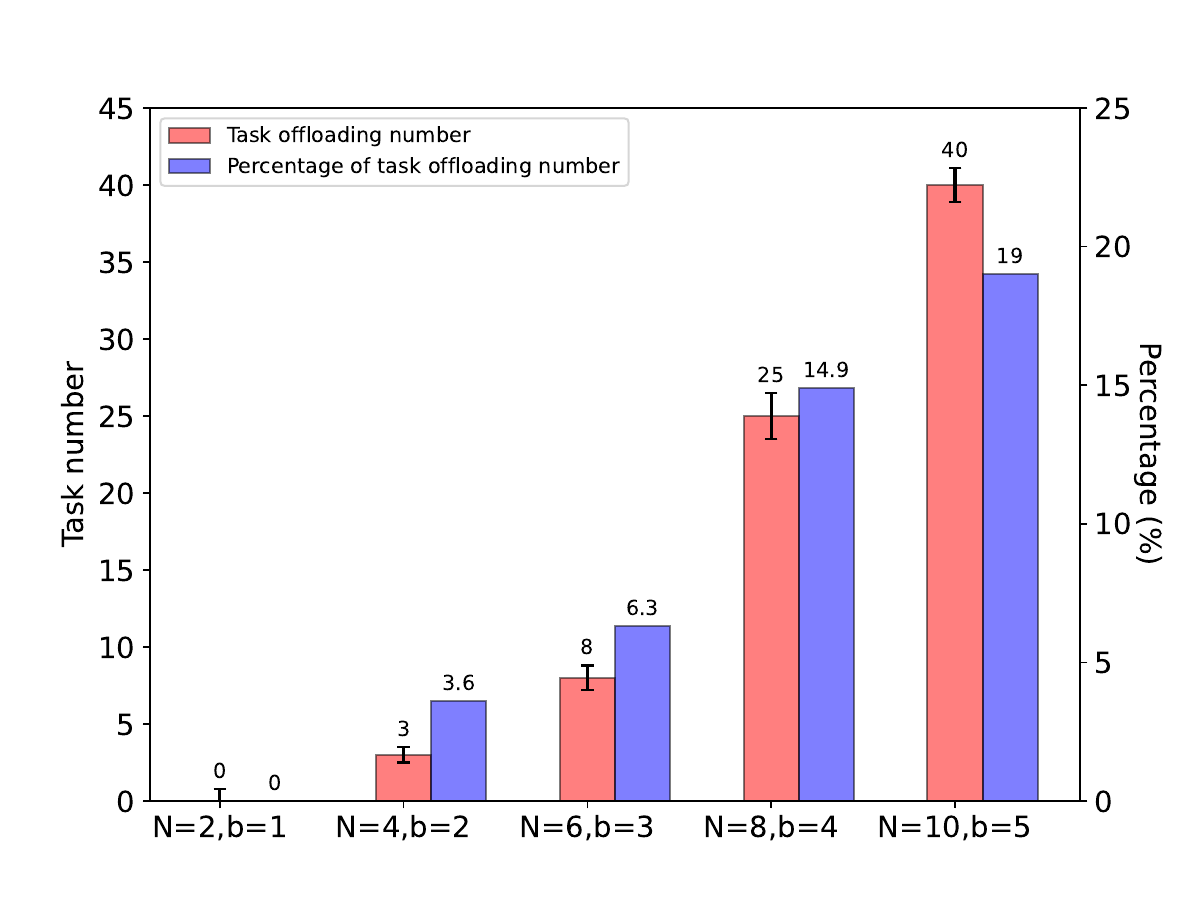}
 \caption{Relationship between workflow scale and the number of vertical offloading tasks}
 \label{task_offloading_number}
\end{figure}

Fig.~\ref{task_offloading_number} depicts the relationship between workflow scale and the number of vertical offloading tasks in five group experiments. Similar to the horizontal roaming experiment, the error bar (standard deviation) of offloading task number changes slightly, which reflects that the Resource Allocator of the workflow scheduling engine has high stability when facing a high workflow workload. As shown in Fig.~\ref{task_offloading_number}, with the increase in the number of injected concurrent workflows, the resources of the cloud-edge cluster become increasingly strained, resulting in the gradual growth of the number of offloading tasks and the percentage of task offloading.

In the $(N=10,b=5)$ group experiment, the number of vertical offloading tasks reaches 40, and the percentage of vertical offloading tasks is $19\%$. In order to verify the ability of the vertical offloading algorithm, we use the resource scaling method in the experiment, and the cloud-edge nodes accommodate more tasks at the same time, causing some task Pods to enter the \verb|OOMKilled| state due to insufficient resources. In the experiment, the cloud-edge workflow scheduling engine can automatically capture the \verb|OOMKilled| task Pod and transfer the load to the corresponding cloud node with the help of the vertical offloading algorithm to realize the resource redistribution of the \verb|OOMKilled| task Pod and ensure the normal execution of the task Pod. 

\textbf{Resource allocation for vertical offloading tasks.} 
We choose task $T7$ of workflow-3 with a vertical offloading event in experiment data group $(N=8,b=4)$ to analyze resource allocation.
At the beginning of the task lifecycle, the cloud-edge workflow scheduling engine also adopts the resource scaling strategy to allocate resources for the current $T7$ as $(CPU:333m,Memory:45Mi)$. Since the allocated memory is less than the minimum amount of memory $Re_{s_{i,j}}^{mem,min}+\beta$ required for the normal execution of task $T7$'s Pod, i.e., 220Mi ($\beta=20Mi$), as shown in Fig.~\ref{task_offloading_task_oom}, task $T7$ encounters \verb|OOMKilled| at the 63rd second.
\begin{figure}[htbp] 
 \centering
 \includegraphics[width=3.3in]{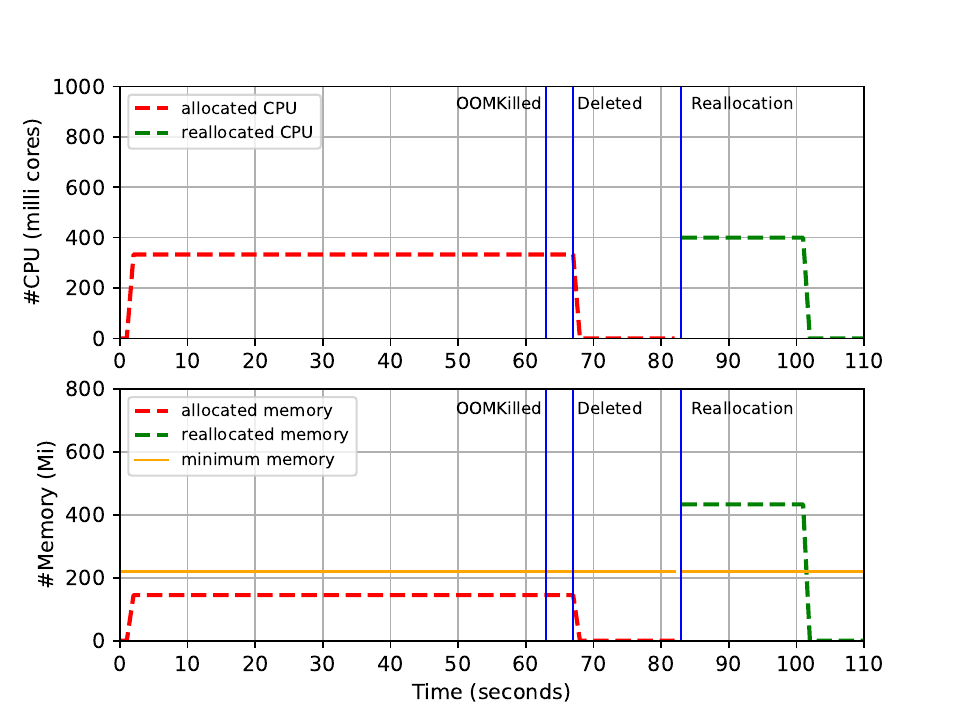}
 \caption{Resource reallocation for vertical offloading task.}
 \label{task_offloading_task_oom}
\end{figure}

$T7$ takes a long time from initial resource allocation to \verb|OOMKilled| (about 62 seconds). It is due to the current cloud nodes node1:192.168.0.161 and node2:192.168.0.162 have fewer resources $(CPU:1000m,Memory:2Gi)$. For the high load of concurrent tasks, the resource scaling method allocates fewer resources (the lower limit of allocated memory resources is set to $100Mi$ in the experiment, to trigger the task Pod \verb|OOMKilled| to verify the task horizontal roaming algorithm and vertical offloading algorithm). When the allocated resource amount is less than the resource lower limit, it will wait until the amount of allocated resources meets the requirement (above the resource lower limit), which consumes a lot of time.
At the 63rd second, the cloud-edge workflow scheduling engine captures this task and calls the task vertical offloading algorithm to remove this task while obtaining the cloud node suitable for vertical offloading of this task. At the 67th second, the Pod corresponding to $T7$ is removed from the cluster. Then, the workflow scheduling engine invokes the resource allocation algorithm at the 83rd second to restart the resource allocation for $T7$. The reallocated resources $(CPU:400m,Memory:433Mi)$ are enough to ensure the smooth operation of $T7$, and finally, at the 98th second, the workflow scheduling engine deletes the successfully executed task $T7$.

\section{Conclusion}
This paper proposes a workflow containerization scheduling scheme KCES for the KubeEdge cloud-edge framework and studies the task scheduling optimization problem of cloud workflow in the cloud-edge collaboration mode. The KCES establishes a workflow scheduling model and cloud-edge node model and uses the resource hybrid management technology to solve the issue of cloud-edge resource collaboration in workflow scheduling. 
Based on the idea of distributed functional roles and the hierarchical division of computing power, the proposed task horizontal roaming and vertical offloading algorithms aim to solve the application collaboration issue between cloud and edge. Experimental results verify the effectiveness of the proposed KCES through the cloud-edge experimental platform. 
In the future, we aim to extend the scale of cloud-edge clusters to integrate reinforcement Learning methods and further improve the performance of KCES.


 




\bibliographystyle{IEEEtran}
\bibliography{IEEEabrv,references}

\end{document}